\documentclass[preprint,3p,12pt]{elsarticle}

\usepackage{amssymb}
\usepackage{placeins}
\usepackage{amsmath}
\usepackage{xcolor}
\biboptions{sort&compress}
\usepackage{setspace}
\usepackage{booktabs}

\journal{Journal TBD}

\begin{document}

\begin{frontmatter}



\title{Deep Learning based Optical Image Super-Resolution via Generative Diffusion Models for Layerwise in-situ LPBF Monitoring}


\author[inst1]{Francis Ogoke}
\author[inst1]{Sumesh Kalambettu Suresh}
\author[inst4]{Jesse Adamczyk}
\author[inst4]{Dan Bolintineanu}
\author[inst4]{Anthony Garland}
\author[inst4]{Michael Heiden}
\author[inst1,inst2,inst3]{Amir Barati Farimani}
\affiliation[inst1]{organization={Mechanical Engineering},
            addressline={Carnegie Mellon University}, 
            city={Pittsburgh},
            postcode={15213}, 
            state={PA},
            country={USA}}

\affiliation[inst2]{organization={Department of Chemical Engineering, Carnegie Mellon University},
            city={Pittsburgh},
            postcode={15213}, 
            state={PA},
            country={USA}}
\affiliation[inst3]{organization={Machine Learning Department, Carnegie Mellon University},
            city={Pittsburgh},
            postcode={15213}, 
            state={PA},
            country={USA}}
\affiliation[inst4]{organization={Sandia National Laboratories},
            city={Alberquerque},
            postcode={15213}, 
            state={NM},
            country={USA}}
\begin{abstract}
The stochastic formation of defects during Laser Powder Bed Fusion (L-PBF) negatively impacts its adoption for high-precision use cases. Optical monitoring techniques can be used to identify defects based on layer-wise imaging, but these methods are difficult to scale to high resolutions due to cost and memory constraints. Therefore, we implement generative deep learning models to link low-cost, low-resolution images of the build plate to detailed high-resolution optical images of the build plate, enabling cost-efficient process monitoring. To do so, a conditional latent probabilistic diffusion model is trained to produce realistic high-resolution images of the build plate from low-resolution webcam images, recovering the distribution of small-scale features and surface roughness. We first evaluate the performance of the model by analyzing the reconstruction quality of the generated images using peak-signal-to-noise-ratio (PSNR), structural similarity index measure (SSIM) and wavelet covariance metrics that describe the preservation of high-frequency information. Additionally, we design a framework based upon the Segment Anything foundation model to recreate the 3D morphology of the printed part and analyze the surface roughness of the reconstructed samples. Finally, we explore the zero-shot generalization capabilities of the implemented framework to other part geometries by creating synthetic low-resolution data. 
\end{abstract}




\end{frontmatter}


\section*{Introduction}
\label{sec:Introduction}

Laser Powder Bed Fusion (LPBF) is a layer-by-layer deposition process, by which a heat source successively melts and fuses thin layers of powder \cite{li2020review, reeves2011additive, beaman2020additive}. While this process enables the creation of parts with small-scale internal features, greater customization and rapid prototyping, its adoption for precision applications is challenged by the occurrence of manufacturing defects at multiple scales \cite{Beese2016, Tofail2018}. These defects often arise from variations in processing parameters during the build, resulting in reduced part quality \cite{debroy2018additive, Morgan2004, Jia2014, Carlton2016}. At the microscale, non-ideal process parameters can introduce porous defects due to the creation of melt pools with non-desired properties. For example, insufficient melting and layer fusion can cause the formation of large, irregularly shaped pores \cite{Sames2014, Darvish2016}. Excessive heat input can produce unstable vaporization cavities that trap gas voids within the solidified part upon collapse \cite{Rehman2021, cunningham2019keyhole}. At the macroscale, geometric defects and inaccuracies can stem from events during the build and powder recoating processes, resulting in a non-uniform powder bed. Foster et al. demonstrate that cantilevered parts with inadequate support undergo deformations, causing unanticipated elevations in the part build \cite{foster2015optical}. These elevations interfere with the operation of the recoater blade, leading to defects \cite{davna2019influence, Kayacan2019}.

Monitoring techniques have been proposed to signal the onset of defects prior to part completion and ex-situ analysis \cite{smith2016spatially, yadroitsev2014selective, berumen2010quality, myers2023high, pak2024thermopore}. Many of these techniques are based upon off-axial sensors, which are easier to retrofit to existing machines, less intrusive than X-ray synchrotron technology, and provide a build-level field of view \cite{abdelrahman2017flaw, li2018situ, foster2015optical}. Specifically, layer-wise optical monitoring enables the detection of in-plane defects, out-of-plane defects, and powder contamination \cite{Mohr2020, pagani2020automated}. For instance, Jacobsmullen et al. leverage optical imaging to analyze part surfaces based on the provided volumetric energy density and detect super-elevation phenomena \cite{zur2013high}. In a related application, Ashby et al. combine thermal simulations with infrared and optical imaging to gain further insight into the surface roughness formation during the printing of overhang features \cite{ashby2022thermal}. Caltanisetta et al. use optical image monitoring to identify a robust method for 3D reconstruction of optically imaged parts, enabling comparisons of their three-dimensional agreements \cite{caltanissetta2018characterization}. High resolutions are required for accurate estimations of the surface roughness in these methods \cite{imani2018process, gobert2018application, SNOW202112}. However, the large resolutions required to detect small-scale features introduce storage constraints during layerwise monitoring \cite{boschetto2024powder}. Additionally, lower-cost imaging systems are more accessible, but fail to maintain the resolution necessary to fully identify features indicating geometric inaccuracies. Therefore, a data-driven model  linking low-cost imaging methods can reduce the overhead costs of layer-wise monitoring, both in terms of integration into existing equipment and storage requirements for real-time monitoring. 

Machine learning tools linking low-cost monitoring data with corresponding high-resolution images provide a pathway towards inexpensive powder bed layer-wise monitoring. This linkage task, known as super-resolution, is well-suited to deep learning models capable of navigating high-dimensional search spaces \cite{li2022srdiff, wang2018esrgan, yang2019deep}. Deterministic models trained on a pixel-wise loss function have been proposed to solve the super-resolution task \cite{zhang2018residual, kim2016accurate}. However, the super-resolution problem is fundamentally ill-posed in a deterministic setting, as multiple feasible high-resolution images can correspond to a single low-resolution image. Specifically, models trained on loss functions that minimize the average error over a set of prediction and target images converge to output an average over the set of feasible target images. This output lacks information about the small-scale structure of the high-resolution image and result in overly-smooth, unrealistic images \cite{li2022srdiff}. Conversely, conditional generative models are trained to model the distribution of target quantities conditioned on an input condition \cite{goodfellow2014generative}, and have been applied in similar contexts within engineering to capture variability \cite{gayon2020pores,  ogoke2022deep}. In this paradigm, a model is trained to capture the probabilistic relationship between high-resolution images and low-resolution images and sample this relationship during inference. This approach retains high-frequency details and preserves realism \cite{wang2018esrgan, rakotonirina2020esrgan+, gao2023implicit}.  Wang et al. demonstrated this paradigm by utilizing a Generative Adversarial Network (GAN) framework to recover high-resolution images from synthetically blurred images \cite{wang2018esrgan}. However, GANs suffer from unstable training and mode collapse effects that impede the capture of an accurate distribution \cite{thanh2020catastrophic, liu2019spectral, srivastava2017veegan}. Similarly, diffusion models have been used to generate high-resolution images from low-resolution inputs.

Addressing the challenges involved in training GAN frameworks, diffusion models divide the generation process into iterative steps to increase stability  \cite{ho2020denoising}, and have also demonstrated promise for super-resolution tasks \cite{li2022srdiff, ogoke2024inexpensive}. However, this division introduces significantly more computational expense during inference \cite{song2020denoising}. Therefore, we adapt the latent diffusion framework to link low-resolution webcam images of the powder bed layer to their high-resolution counterparts \cite{Rombach_2022_CVPR}. Encoding the images into a latent space prior to the diffusion process reduces the time and memory required for inference, a critical consideration for in-situ monitoring. We evaluate the performance of our implemented framework based on the geometrical agreement between the predicted 3D part morphology and the high-resolution 3D morphology. Additionally, we investigate the ability of the model to recapture the surface roughness of the constructed parts. Finally, we investigate the performance of the model on transfer learning tasks between part builds using synthetically generated low-resolution data. 

\section*{Methods}
\label{sec:Methods}

\subsection*{Diffusion Implementation}

\begin{figure}[hbt!]

\centering
\includegraphics[width=1\linewidth]{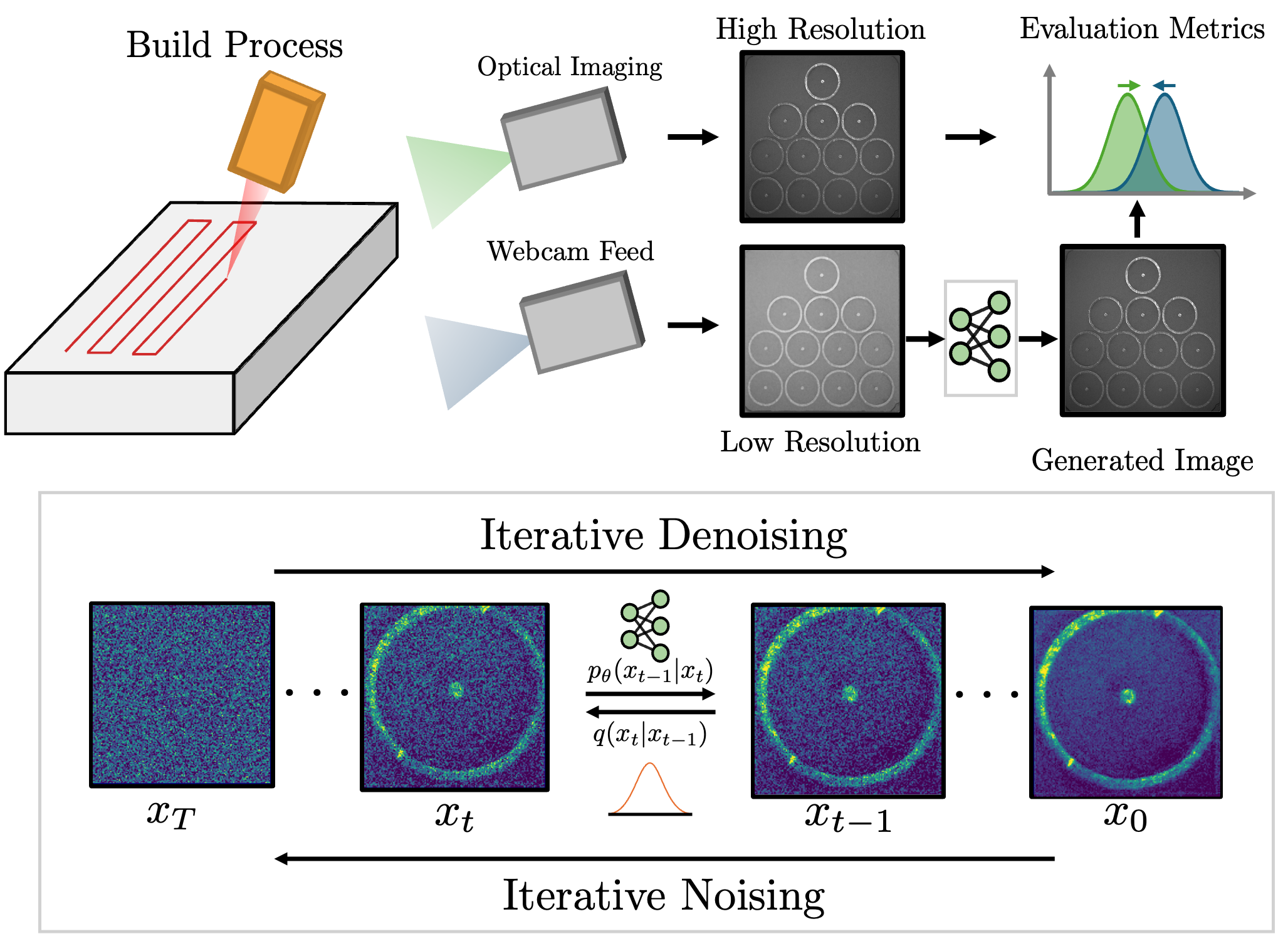}

\caption{\textbf{a)} A schematic of the proposed workflow. During the build process, layerwise high-resolution optical images and low-resolution webcam images are collected. The low resolution images undergo image translation to the space of high-resolution optical images via a generative model. The performance of the upscaling process is evaluated based on distributional comparisons between the high-resolution ground truth and the generated high-resolution images.\textbf{ b)} A diffusion model is used to generate new high-resolution samples conditioned on the input low-resolution webcam images, through an iterative denoising training process.
 }
\label{fig:overall_schematic}
\end{figure}

We apply a Denoising Diffusion Probabilistic Model (DDPM) to stochastically link the low-resolution monitoring images to the high-resolution monitoring images \cite{ho2020denoising}. DDPMs are Markov chain-based latent variable models that learn to produce synthetic samples covering the complex empirical distribution of the dataset. Specifically, a DDPM is trained with variational inference to iteratively transform a standard Gaussian distribution into an arbitrarily complex distribution defining the observed samples. In the unconditional case, this is equivalent to learning to model and sample from the empirical distribution $p(\mathbf{x})$, where $\mathbf{x}$ is the set of I.I.D. data samples. During this process, a sample from a Gaussian distribution is iteratively transformed through successive denoising operations until it acts as a sample from the complex data distribution that models the empirical behavior of the dataset. In the conditional generation case, the task is modified to model the conditional probability of observing a specific random variable $\mathbf{x_0}$ given prior knowledge $y$. This requires the model to learn the empirical data distribution $p (\mathbf{x} | y)$. 

To model this empirical data distribution, the diffusion process is divided into two stages centered around the concept of denoising, namely, the forward stage where the model is trained, and the backward process, where the trained model is queried to produce individual samples. During the forward process, noise is added to a ground truth image, $x_0$, from the dataset over $T$ time steps, transforming the initial sample into isotropic Gaussian noise. At each stage of this forward process, the model is trained to predict the amount of noise added to the ground truth image from an arbitrarily noisy version of this image, $x_t$.

\begin{figure}[hbt!]

\centering
\includegraphics[width=1\linewidth]{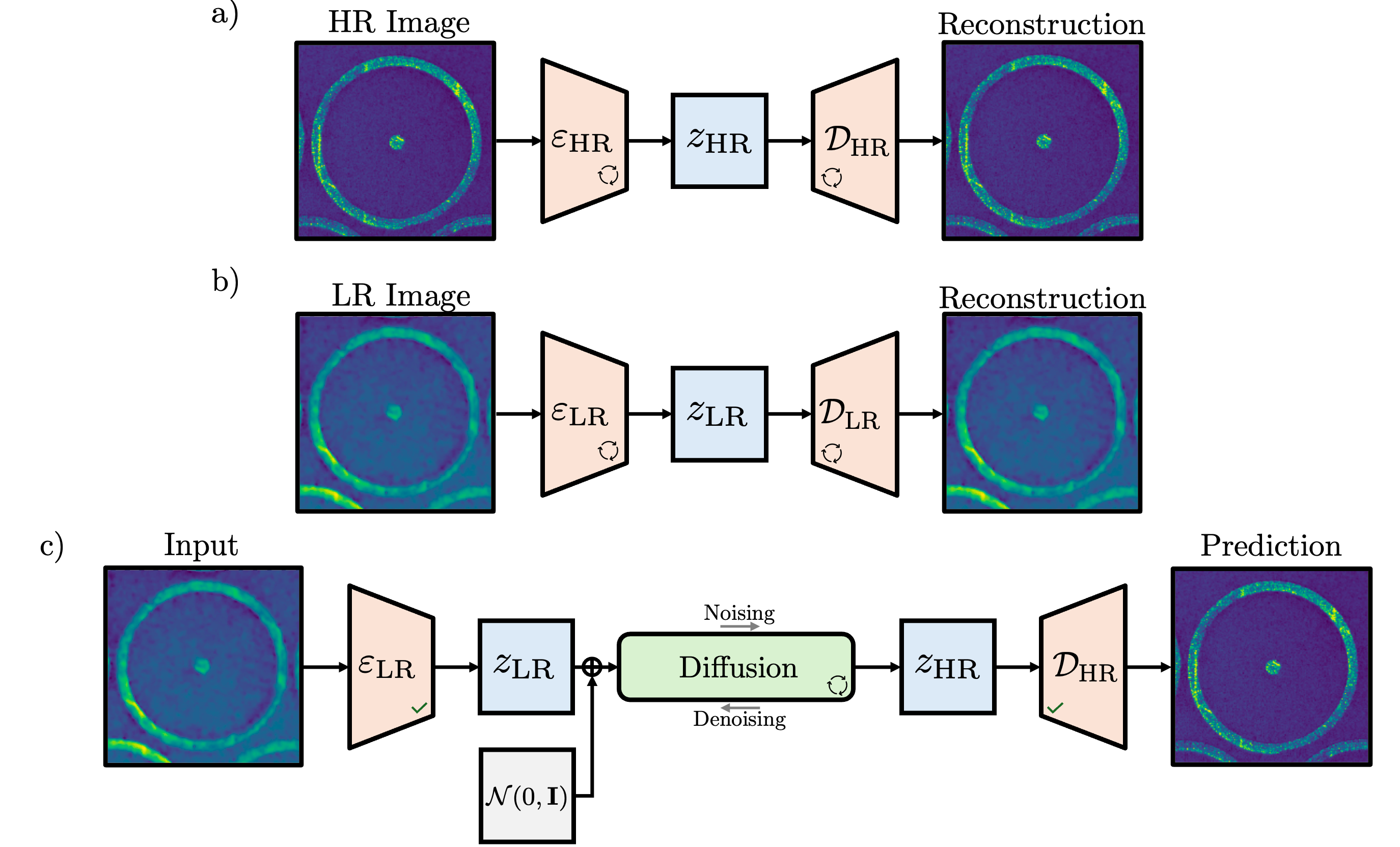}

\caption{\textbf{a), b)} Two autoencoder networks are trained to encode the layerwise patches for each patch-wise image in the dataset to a latent space, $z$. One encoder is trained to learn an embedding of the high-resolution (HR) data, and a second encoder is trained to learn an embedding of the low-resolution (LR) input data. \textbf{c)} During the diffusion model training process, the trained autoencoders are used to first project the low-resolution data into a compressed latent space. Next, a conditional diffusion model generates an appropriate high-resolution latent vector from the low-resolution input data. The high-resolution decoder network is then used to reconstruct a predicted high-resolution sample.}
\label{fig:ae_schematic}
\end{figure}

\begin{figure}[hbt!]

\centering
\includegraphics[width=1\linewidth]{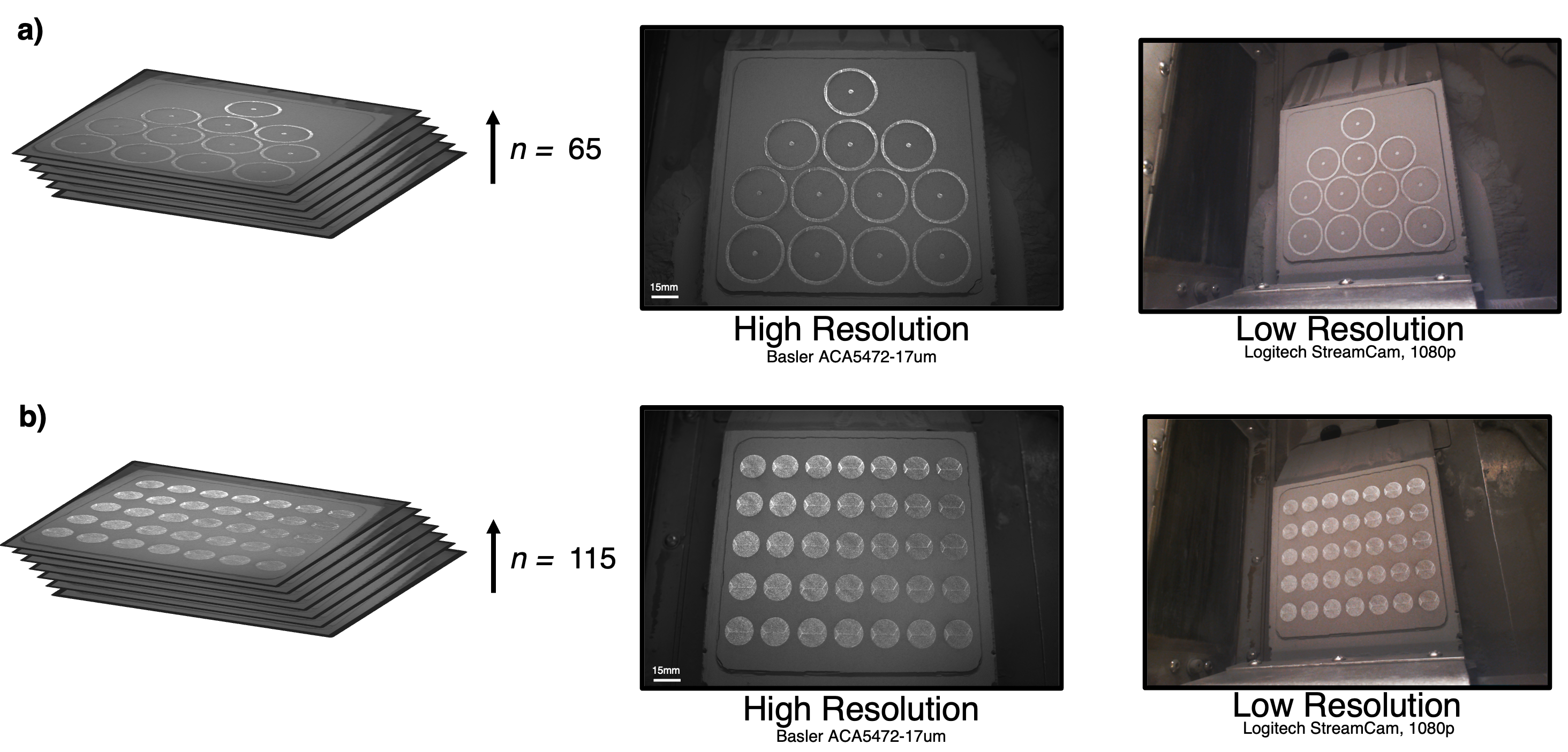}

\caption{Layer-wise images are collected during the build process for two sample groups of parts. These images are collected in both high resolution with a Basler AC45472-17um camera and low resolution with a 1080p Logitech StreamCam camera. \textbf{a)} $n$ = 65 layer-wise images are collected for the first build. \textbf{b)} $n$ = 115 layer-wise images are collected for the second build.}
\label{fig:dataset_details}
\end{figure}

\begin{figure}[hbt!]

\centering
\includegraphics[width=1\linewidth]{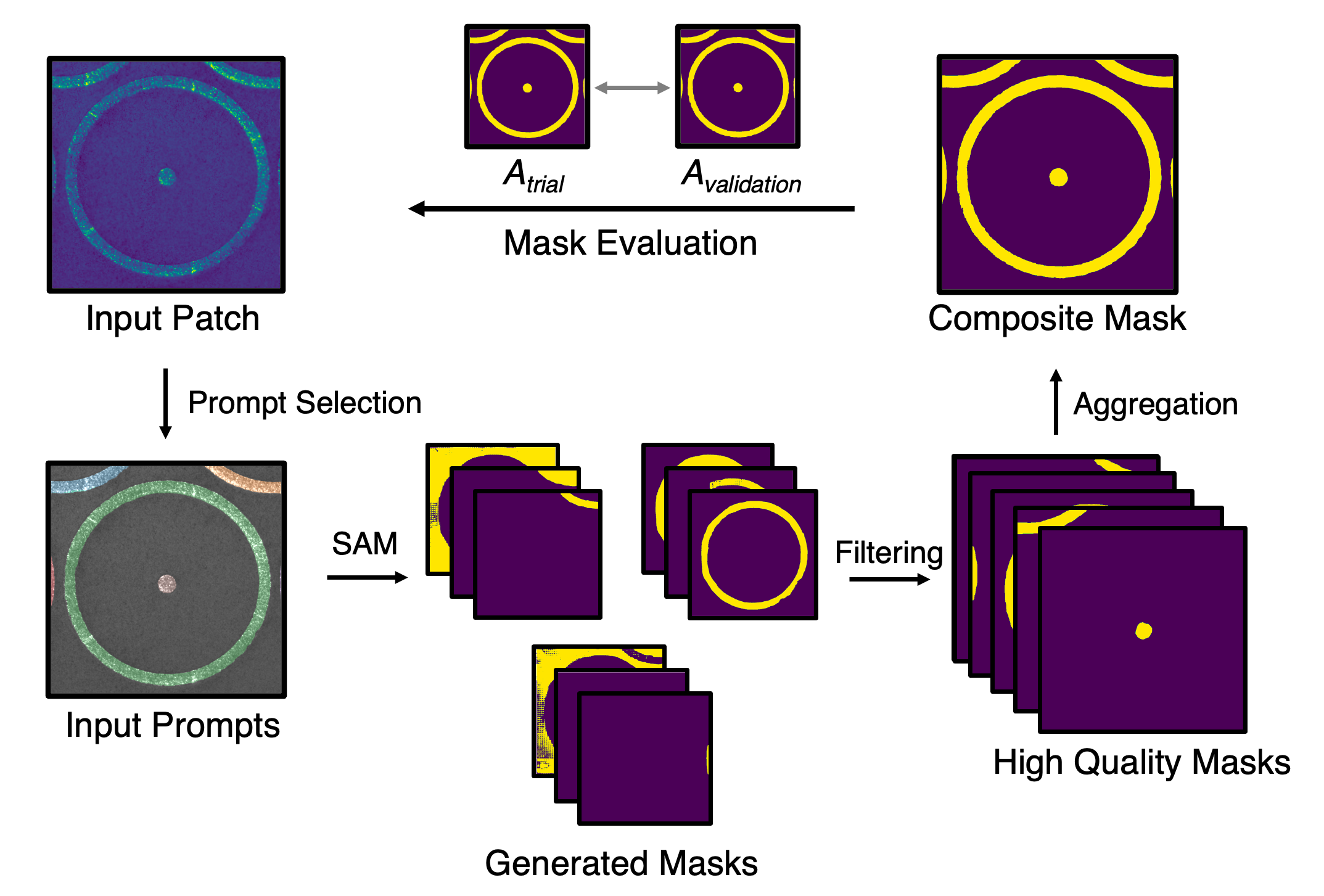}

\caption{A two-stage process is used to automatically segment arbitrary patches of the build plate with the pre-trained Segment Anything foundation model. In the first stage, an adaptive threshold is applied to extract areas of the part surface, which are labeled to indicate the visible part components. In the second stage, query points are sampled for use with Segment-Anything, which provides a series of part masks. These part masks are aggregated to form the final composite mask.}
\label{fig:segmentation_architecture}
\end{figure}

\begin{figure}[hbt!]

\centering
\includegraphics[width=1\linewidth]{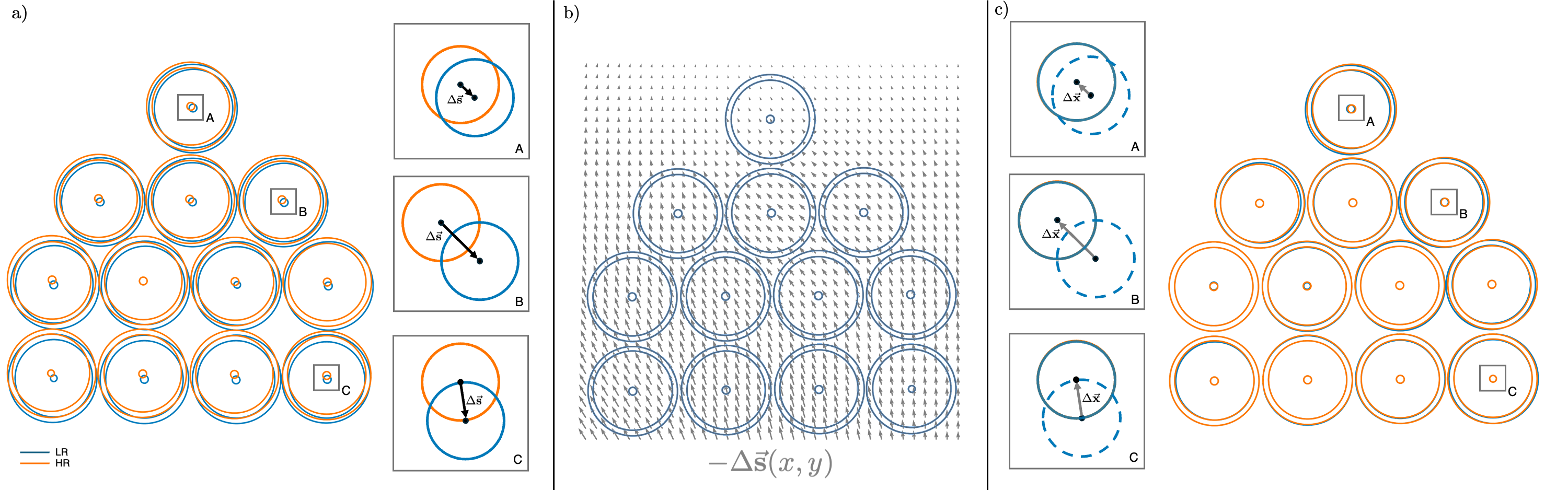}

\caption{Layer-wise registration between the high-resolution and low-resolution images obtained at different orientations within the build chamber, the two image fidelities are registered to each other following an initial warping transformation. \textbf{a)} A comparison of the overlaid part samples after an initial warp to approximate the perspective transformation from the original point-of-view to a top-down view. The displacement between each part across the two image fidelities are calculated as $\Delta \mathbf{s}$. \textbf{b)} The measured displacements between the parts are interpolated across the build plate to form a deformation map, $-\Delta \mathbf{s}(x,y)$. \textbf{c)} The high-resolution image is transformed according to the deformation map  $-\Delta \mathbf{s}$ to achieve pixel-wise agreement.}
\label{fig:deformation_map}
\end{figure}

During the backward process, the trained model is used to perform inference by transforming a random normal Gaussian sample to a sample drawn from the empirical data distribution. To do so, an isotropic Gaussian sample is drawn from the normal distribution, and denoised over $T$ timesteps by using the trained model $p_\theta(x)$ to predict the amount of noise to remove at each timestep. The scale of the noise added at each timestep is determined by the variance schedule, which is designed to increase at the stages closest to the isotropic noise distribution, and decrease as the samples become closer to a high resolution sample. A U-Net architecture is used to parameterize $p_\theta$ \cite{ronneberger2015u}.

\subsection*{Latent Diffusion}

Due to the memory and time constraints involved in performing multiple diffusion steps during inference, the sample generation process is conducted in a learned latent space. To construct this latent space, two autoencoder networks are trained, one encoder network compressing the high-resolution data, and one encoder network compressing the low-resolution data. Each autoencoder network compresses either the high-resolution or low-resolution modes of data to define an efficient latent space that preserves the important details of the input image while neglecting imperceptible details. In the original paper by Rombach et al. \cite{Rombach_2022_CVPR}, the low-resolution image is provided directly to the diffusion model as the conditioning input. However, in our application, the high-resolution samples are first downscaled to the same image size as the low-resolution samples to reduce the memory requirements of the machine learning model. This is done because much of the information lost during the transition from Basler optical imaging to  webcam imaging is due to inability of the camera sensor to detect fine detail, as opposed to the pixel resolution itself being insufficient. Therefore, by framing this task as an image translation task between low-detail and high-detail images at the same pixel resolution, we reduce the amount of computing power that must be dedicated to the latent diffusion process significantly. To account for this modification, the encoder network described earlier compresses the low-resolution webcam images to a smaller two-dimensional latent space of identical size to the high-resolution latent space.

Once these autoencoder networks are trained (Figure \ref{fig:ae_schematic} a), b)), their frozen model weights are used to compute a latent space for the diffusion process. Specifically, the low-resolution input $x_{LR}$ is first transformed to a latent vector $z_{LR} = \epsilon_{LR} (x_{LR})$ , and concatenated to the noised sample $x_t$ at each timestep of the forward process. From this concatenated input, the model predicts the noise $\epsilon$ added to the data sample $x_0$ to produce $x_t$. In the reverse process of the latent diffusion model, an initial sample from the standard normal Gaussian distribution $x_T$ is iteratively denoised by removing the predicted noise added at each timestep $\epsilon_\theta(z_{LR}, t)$ from $t = T$ to $t = 1$. At the conclusion of the denoising process, $x_0$ is a prediction of the latent vector $z_{HR}$. This latent vector $z_{HR}$ is decoded by the high-resolution decoder network to predict the equivalent corresponding high-resolution image of the build plate.

By leveraging this latent diffusion framework, a greater number of samples can be generated in parallel to reduce the time required for sample generation.

\subsection*{Image Segmentation}

Layer-wise optical images can yield 3D estimates of the part structure by segmenting each sample into a part phase and powder bed phase, and stacking the segmented part phase masks for each layer. To perform the segmentation task, we use a pre-trained segmentation foundation model, Segment Anything \cite{kirillov2023segment}. Segment Anything is a segmentation model designed to predict object masks from arbitrary images, and trained on 1 $\times 10^9$ samples and masks of images for generalizable performance. For each application of Segment Anything (SAM), a query prompt consisting of specific points labeling the foreground regions of interest and exclusionary points labeling the background regions are provided to the model.  Following this, multiple masks are produced for each query point provided. To ensure reliable mask detection for high-throughput scenarios, we design a two stage process for creating a reliable composite mask from a sequence of auto-extracted query points.  During the initial object detection stage, we implement a workflow to automatically extract the relevant query points in a consistent manner before using they are used as SAM prompts.

An object detection method is first applied to detect unique potential part components on the build plate. To do so, an adaptive Sauvola binarization threshold is first applied to generate an initial estimate of the pixels belonging to a foreground structure in the analyzed image. Due to the intrinsic noise present within the image, these initially extracted foreground masks are often disconnected and sparse. Therefore, a dilation operation is applied to morphologically close small holes within the image to form contiguous structures for each part component visible in the image. Following this, a connected components algorithm is used to distinctly label each of the part components present within the build plate patch. Finally, a set of $m+1$ query points is sampled for each part component identified. One query point is selected at the point where the Euclidean distance transformation between interior edges of the part component is maximized, while $m$ additional query points that coincide with the initial mask are also selected at random locations coinciding with the identified part component.

Following the mask-filtering described in the auto mask generation  process in \cite{kirillov2023segment}, all $3(m+1)$ estimated SAM masks undergo a series of filtering operations to remove low-quality masks and deduplicate identified regions. Specifically, masks which do not meet the thresholded confidence and stability scores are no longer considered, in addition to masks that duplicate others while having a lower estimated confidence score. This process results in a list of independent high-confidence masks, which are aggregated together via a union operation. However, unexpected errors in the object detection process may result in background regions of the image being incorrectly labeled as foreground prompts. To resolve this, we implement an adaptive prompt modification process that iteratively compares the area of the final segmented image to a manually inspected ground truth mask to ensure agreement. Specifically, we exploit the fact that each patch at a given layer will have the same field of view, and calculate the area of the segmented high-resolution image corresponding to that patch. Based upon this ground truth area, we define two scenarios to develop a heuristic for optimizing the placement of the query points based upon the calculated area of the trial mask. If the area of the mask is smaller than 80 \% of the ground truth area, the mask is identified to have missed at least one part object visible in the image, and the number of query points per object $m$ is increased. If the area of the mask is larger than 120 \% of the ground truth area, it suggests that an area of the background is falsely segmented as the foreground. In this scenario, the number of additional points $m$ sampled for each object is set to 0, and an additional exclusionary prompt point is sampled within the background to ensure the background is correctly segmented. This process repeats for up to 10 iterations until convergence. If no viable mask is identified after 10 iterations, the layer is denoted as an anomaly and excluded from the downstream analysis of the three-dimensional part morphology.

\subsection*{Data Generation}

 A NI CompactDAQ loaded with 3 NI-9223 voltage data acquisition cards was used to record process data from the Galvonometer controller on a 3D Systems ProX DMP200 laser powder bed fusion additive manufacturing system. A low-resolution webcam (Logitech StreamCam, 1080p) and a high resolution scientific camera (Basler ACA5472-17um with Edmund Optics 86-71 Lens) were installed inside of the build chamber of the ProX 200. The cameras were mounted off-axis from the laser, pointing at the build plate from different angles. Image acquisition was performed using a Python script along with the OpenCV-Python and PyPylon libraries. The Python script waits for the laser to turn off before capturing an image in order to ensure each layer was complete before images were captured. Images of the completed layer from both the low resolution and high resolutions cameras were then automatically written to a disk with the filename corresponding to the layer number. AISI 316 L stainless steel powder (3D Systems) was used to fabricate small ring-like parts on the ProX DMP200 machine with a continuous Yb-fiber laser (1070 nm wavelength). Nominal settings included a laser hatch spacing of 50 $\mu$m and a laser focus positioned 1.5 mm below the powder surface. Parts were printed with a “hexagons” scan pattern (standard for 3D Systems), where the laser raster scans back and forth within 10 mm diameter circumscribed hexagon islands, stitched together to generate part layers. During this study, the actual laser power was 113 W and laser scan speed was 1400 mm/s. Oxygen content in the build chamber was kept below 1000 ppm and a constant argon cover gas flow was maintained across the build area during the process.

\subsection*{Data Processing and Alignment}

The webcam images and the Basler camera images are taken at different orientations, requiring warping to achieve pixel-wise alignment. This pixel-wise alignment requires the identification of each part on the build plate. The automation of this decomposition process requires a method to identify the part object location on the build plate. To achieve this, we first apply a Hough transform to detect the circular features of each of the parts on the build plate and then compute a bounding box which inscribes each part.

Once the relative position of each individual part has been identified on the build plate, we aim to achieve  pixel-level correspondence between the low-resolution and high-resolution samples. This alignment is crucial for downstream machine learning predictions that use smaller segments of the build plate image for training. For this alignment process, we construct a deformation map between the low-resolution and high-resolution samples. This is done by identifying the center of each part for the low-resolution and high-resolution samples based on the Hough transform estimates. The distance between corresponding part centers in the low-resolution and  high-resolution images are interpolated to create a 2-D deformation map, denoting the correct $x$ and $y$ shifts necessary to achieve pixel-level correspondence between the webcam images and Basler camera images. The high-resolution image is then warped to reverse this computed deformation map and ensure alignment.

\section*{Results}
\label{sec:Results}

\subsubsection*{Metrics}

We define several metrics to evaluate the performance of the generative model in reconstructing the high-resolution features visible from the Basler camera. Specifically, we compute conventional reconstruction based metrics, such as the mean absolute error (MAE) of the pixel intensities predicted by our model. We also compute image processing based metrics for evaluating model performance, including the peak signal-to-noise ratio (PSNR), and the structural similarity index measure (SSIM).

The peak signal-to-noise ratio is given by the expression in Equation \ref{eq:psnr}, and measures the ratio of the preserved signal in a reconstructed sample to the amount of signal-distorting noise present in the sample.
\begin{equation}
\label{eq:psnr}
\mathrm{PSNR} = 10 \cdot \log_{10} \left( \frac{\mathrm{MAX}_I^2}{\mathrm{MSE}} \right)
\end{equation}

The structural similarity index measure (SSIM) describes the interdependence of pixels that are in close spatial proximity in the image to estimate the perceived change in image structure \cite{wang2004image}. The SSIM between a pair of images ($I_x$, $I_y$) is given by Equation \ref{eq:ssim}, calculated over individual windows of each image $x$ and $y$. In Equation \ref{eq:ssim}, $\mu_x$ is the pixel-wise mean of window $x$, $\mu_y$ is the pixel-wise mean of window $y$, $\sigma_x^2$ and $\sigma_y^2$ are the pixel-wise standard deviation of each window respectively, and $C_1$ and $C_2$ are constants derived from the dynamic range of the image to avoid division by zero.
\begin{equation}
\label{eq:ssim}
\mathrm{SSIM}(x, y) = \frac{(2\mu_x\mu_y + C_1)(2\sigma_{xy} + C_2)}{(\mu_x^2 + \mu_y^2 + C_1)(\sigma_x^2 + \sigma_y^2 + C_2)}
\end{equation}

To analyze the performance of the model in reconstructing the image texture present in the high-resolution image, we apply complex wavelet transforms to analyze the statistics of the two-dimensional random process representing the powder bed. Specifically, we implement the loss function proposed in \cite{zhang2021maximum} for an image generation task to benchmark the performance of the predicted samples. This approach is based on the fact that the covariance of a phase harmonic operator applied to a wavelet transform encodes information about the non-linear dependencies across frequency and scale. The scale-dependent variation of these complex wavelet transforms have been shown to provide information about the geometrical structures present within an image \cite{grossmann1990reading}. Following the approach detailed in \cite{zhang2021maximum}, we apply complex bump-steerable wavelet transforms to encode the wavelet coefficients across multiple scales and angles. For consistency, both the high-resolution and low-resolution samples are upscaled to 512 $\times$ 512 prior to the phase harmonic analysis. The covariance distance (CVD) loss metric used to quantify the image texture is 

\begin{equation}
\label{eq:loss}
    CVD = \frac{\| \tilde{K}_{\mathcal{R}x} - \tilde{K}_{\mathcal{R}\bar{x}} \|}{ \| \tilde{K}_{\mathcal{R}\bar{x}} \|  }
\end{equation}

where $ \tilde{K}_{\mathcal{R}x}$ is given by Equation \ref{eq:covariance}, where $\mathcal{R}_v$ are the wavelet transform coefficients extracted at a given spatial location, scale, and angle and  $\mathcal{R}_v'$ are the wavelet transform coefficients extracted at a neighboring spatial location, scale, and angle. 

\begin{equation}
\label{eq:covariance}
   \tilde{K}_{\mathcal{R}x} = \frac{1 }{|G|} \sum_{g \in G} \left ( \mathcal{R}_v (g(x) - \tilde{M}_{\mathcal{R}}(v) 
\right ) \left ( \mathcal{R}_v (g(x) - \tilde{M}_{\mathcal{R}}(v') 
\right ) 
\end{equation}

$G$ defines a set of translation operations designed to induce translational invariance, where $g$ is an individual translation operation \cite{zhang2021maximum}. Additional detail regarding the calculation of $\mathcal{R}(x)$ is provided in \ref{sec:wavelets}.

\subsection*{Patch-based Training}

\begin{figure}[hbt!]

\centering
\includegraphics[width=1\linewidth]{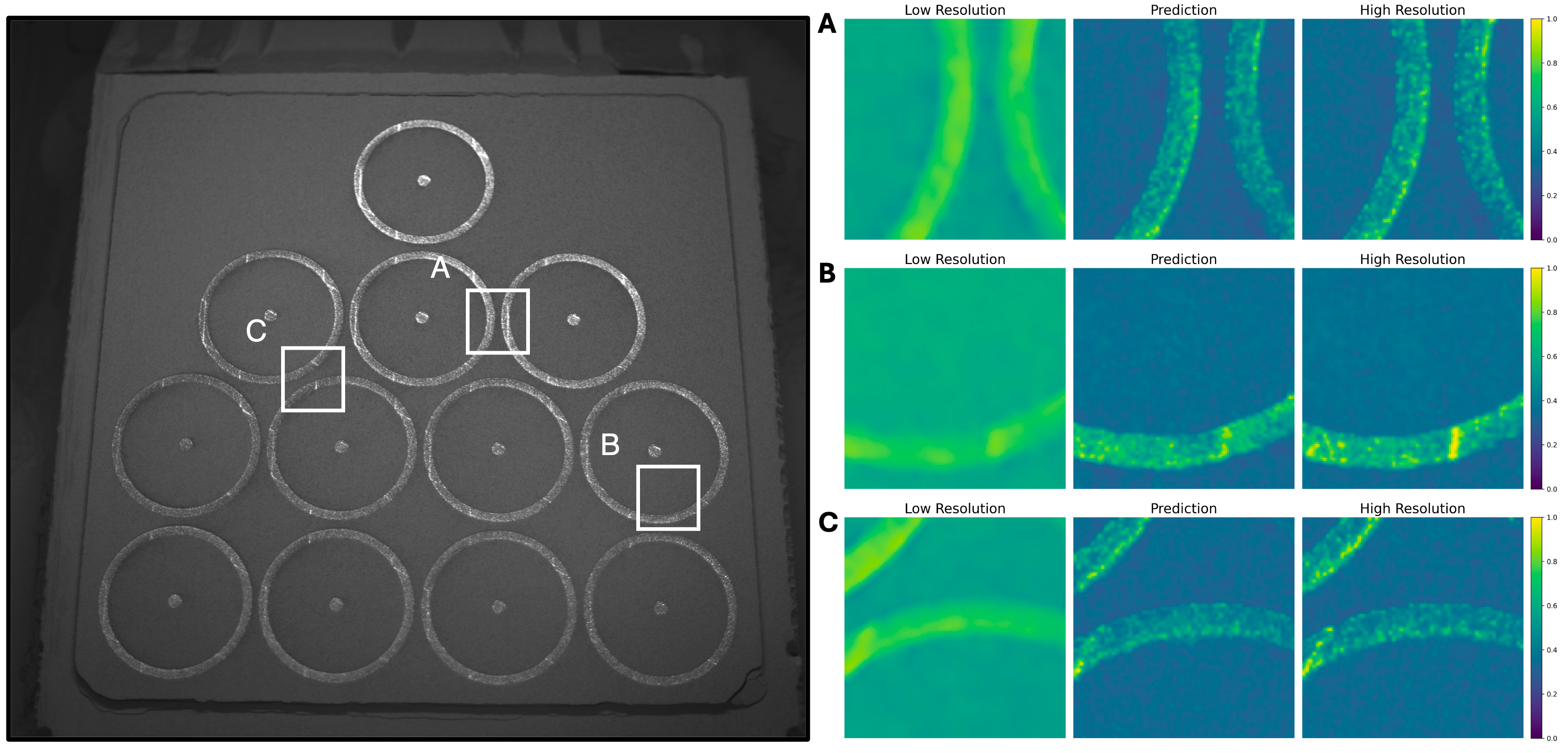}

\caption{Patches are extracted from each layerwise image of the build plate and used as input to the super-resolution model. (Left) A sample layer-wise image, with three example patches highlighted in white. (Right) The low-resolution, high-resolution, and predicted patch-wise images for each of the three patches highlighted in the left panel.}
\label{fig:patch_comparison}
\end{figure}
The first experiment conducted extracts individual 64 $\times$ 64 size patches from the layer-wise build plate in order to upscale them to the corresponding high-resolution patch images. With this configuration of the model training process, we aim to efficiently upscale the entire build plate in a piecewise manner, avoiding the memory constraints of processing a build-plate level image.  Each autoencoder model is trained for 100 epochs with a KL-divergence loss, perceptual loss and discriminator loss with a learning rate of $4.6 \times 10^{-5}$. Following this, the latent diffusion model is trained for 300 epochs with a learning rate of $1 \times 10^{-5}$. After applying a random patch sampling on each dataset, 115 patches are extracted for each layer. An 80\%-20\% train/test split is applied, where 80\% of the layers are seen during training.

The qualitative performance of the patch-based upscaling task is demonstrated in Figure \ref{fig:patch_comparison} for three sample patches on the build plate. For each patch, the low-resolution sample approximately captures the large scale structure of the part, but fails to capture details regarding the powder bed surface. However, the predictions obtained by the super-resolution model are able to reconstruct the variation observed within the powder bed alongside the sharp demarcation between the powder bed and the part cross-section. This is key for potentially detecting defects such as recoater impact, spatter and uneven spreading that deposit residue on the powder bed.

\begin{table}[hbtp!]
\centering
\begin{tabular}{@{}llcccc@{}}
\toprule
\multicolumn{2}{c}{Configuration} & \multicolumn{4}{c}{Image Metrics}
\\ \cmidrule(lr){1-2} \cmidrule(lr){3-6}
Dataset   & Comparison & MAE  $\downarrow$   & PSNR $\uparrow$  & SSIM $\uparrow$   & CVD  $\downarrow$  \\ \midrule
Dataset A & HR $|$ LR    &  0.241   &  3.20   &     0.512 &    3.03 $\times 10^6$      \\
Dataset A & HR $|$ SR    & \textbf{0.017 } & \textbf{23.3 } & \textbf{0.524}        & \textbf{1.45} $\mathbf{\times 10^5}$  \\ \midrule
Dataset B & HR $|$ LR    & 0.134 &  15.0  & \textbf{0.511} & 1.31 $\times 10^4$   \\
Dataset B & HR $|$ SR     &  \textbf{ 0.043} &  \textbf{21.4} & 0.464 & \textbf{1.11} $\mathbf{\times 10^3}$ \\ \bottomrule
\end{tabular}
\caption{A comparison of the dataset metrics defining the ability of the model to perform image reconstruction and part reconstruction.}
\label{tab:patch_comparison}
\end{table}

\begin{figure}[hbt!]

\centering
\includegraphics[width=1\linewidth]{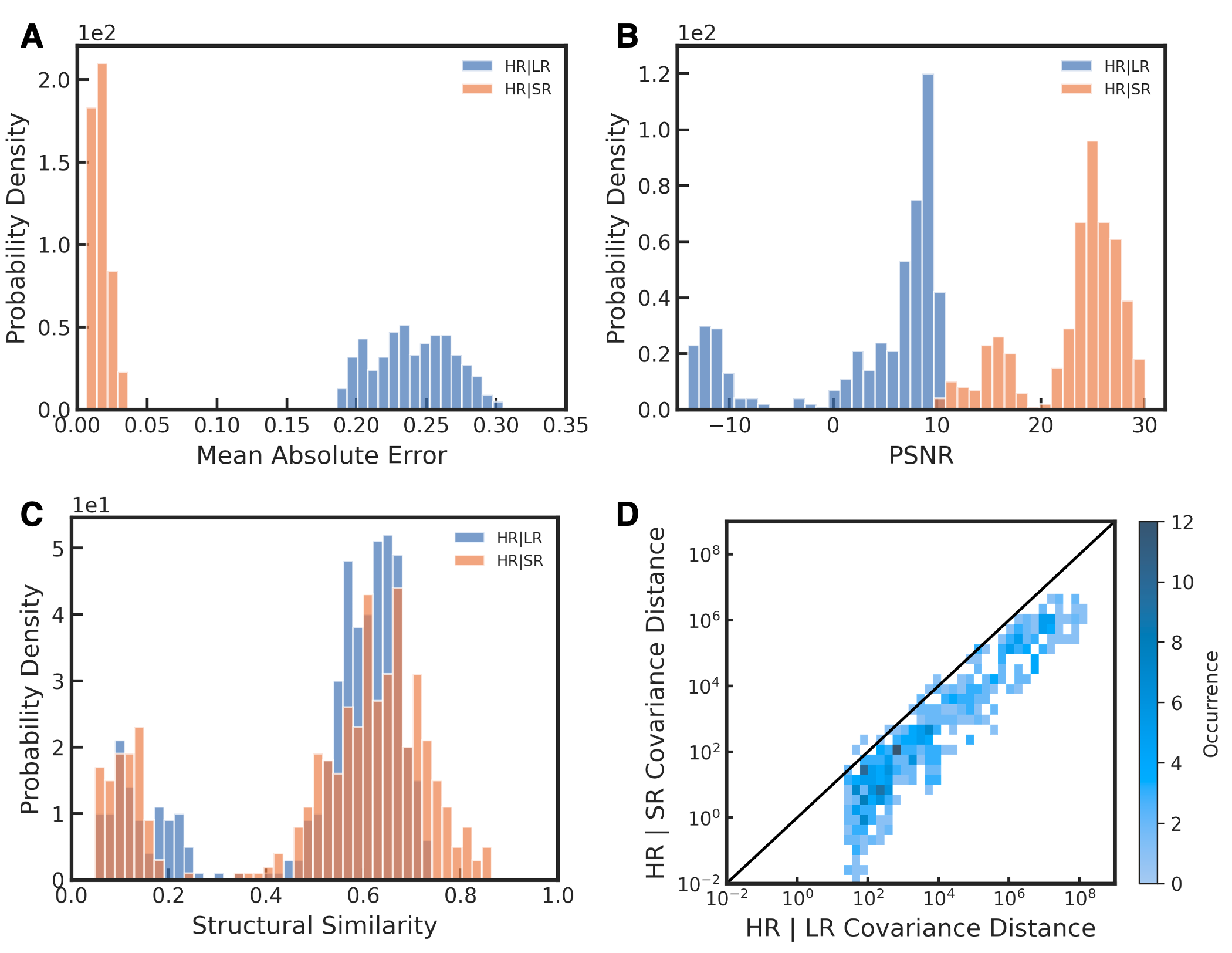}

\caption{Distributional comparisons of the image-based metrics studied to quantify the performance of the super-resolution task. For each metric, the agreement between the high-resolution and low-resolution sample (HR-LR) is compared to the agreement between the high-resolution and predicted upscaled low-resolution sample (HR-SR). \textbf{a)} A distributional comparison of the peak-signal-to-noise ratio (PSNR). \textbf{b)} A distributional comparison of the structural similarity (SSIM). \textbf{c)} A distributional comparison of the Mean Absolute Error (MAE). \textbf{d)} A distributional comparison of the covariance distance (CVD).} 
\label{fig:histogram_figure_a}
\end{figure}

\begin{figure}[h!]

\centering
\includegraphics[width=1\linewidth]{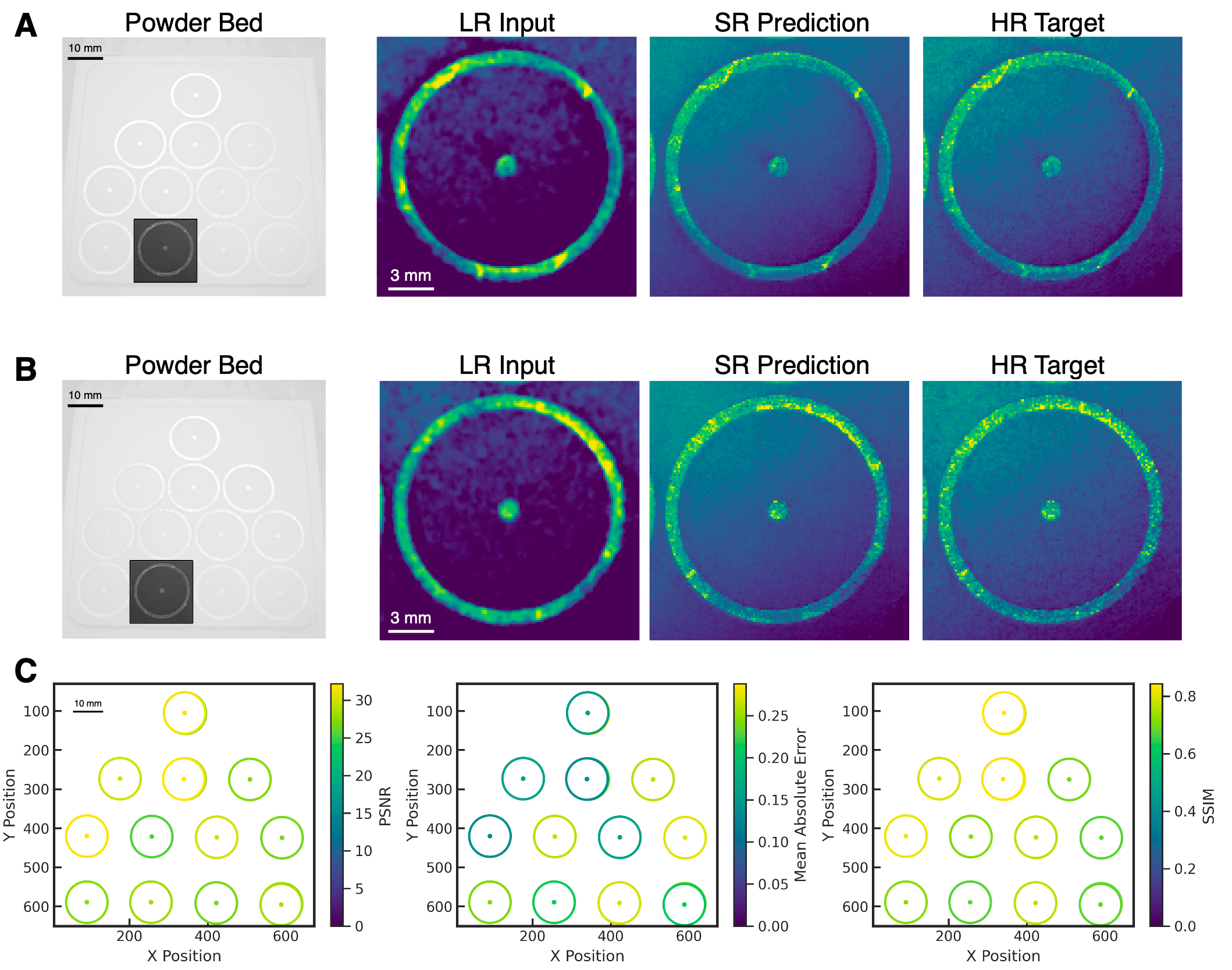}

\caption{Bounding boxes enclosing individual parts on the powder bed are extracted for further analysis of the part reconstruction capabilities of the model. \textbf{a), b)} The low-resolution input, super-resolution model prediction, and high-resolution target image for two sample parts on the build plate. \textbf{c)} Three of the image metrics studied, plotted for each sample part to show the spatial variation of the image reconstruction quality based on image lighting. The highest PSNR, highest SSIM, and lowest MAE are achieved in the brighter sections of the build plate.}
\label{fig:part_split_qualitative}
\end{figure}

The quantitative agreement between the ground truth patch structures and the predicted patch structures is shown in Figure \ref{fig:histogram_figure_a}. To do so, the PSNR, SSIM, and MAE are extracted to compare the deviation between the high-resolution and low-resolution samples (HR $|$ LR) to the deviation between the high-resolution and predicted samples (HR $|$ SR). Notably, a higher PSNR and lower MAE is observed uniformly for the latent diffusion predictions, indicating significant improvements in overall image quality from the low-resolution sample. A higher SSIM is also observed on average for the model predictions, though this effect is less pronounced due to the existing large scale structural agreement between the high-resolution and low-resolution samples. The covariance distance also decreases when compared to the low-resolution image, indicating that the latent diffusion model is able to correctly predict the high-frequency structure present within the Basler camera image.

\begin{table}[ht]
    \caption{Performance Metrics for the Latent Diffusion and Pixel Diffusion models.}
    \centering
    \label{tab:model_performance}
    \begin{tabular}{lcccccc}
        \toprule
        Model & PSNR $\uparrow$ & MAE $\downarrow$ & nCVD $\downarrow$ & SSIM $\uparrow$ & Inference Time (s) $\downarrow$\\
        \midrule
        Latent Diffusion & \textbf{23 $\pm$ 4.7} &\textbf{ 0.017 $\pm$ 0.01} & 1.9 $\pm$ 1.1 & \textbf{0.52 $\pm$ 0.23} & \textbf{0.01} \\
        Pixel Space Diffusion & 21 $\pm$ 4.6 & 0.025 $\pm$ 0.01 &\textbf{ 0.28 $\pm$ 0.18} & 0.46 $\pm$ 0.21 & 0.13 \\
        \bottomrule
    \end{tabular}
    
\end{table}

Finally, we compare the performance of the latent diffusion model to the performance of a pixel-space diffusion model, with the hyperparameters held constant. Notably, the PSNR, MAE, and SSIM values remain consistent between the more computationally expensive pixel diffusion model, and the implemented latent diffusion model (Table \ref{tab:model_performance}). Due to the diffusion process occurring in a reduced order $16 \times 16$ latent space, we achieve inference in a time of 0.01 s/sample at a batch size of 500, compared to the 0.13 s/sample inference time for the pixel space diffusion model.


\subsection*{Part-based Training}

Next, we aim to examine the performance of the model towards correctly resolving the three-dimensional structure of the part after segmentation has been applied. To avoid data leakage between the train-test set and to ensure each part in the test set has not previously been seen during training, the train-test split is modified to take place in the space of parts on the build plate. Specifically, for each layer 80\% of the parts on the build plate are taken for training, while the remainder 20\% are used to evaluate generalization performance to unseen samples. In this experiment, the model hyperparameters are held constant from the patch-based training case. Each part sample is resized to a 128 $\times$ 128 image prior to the training process.

\begin{figure}[h!]

\centering
\includegraphics[width=1\linewidth]{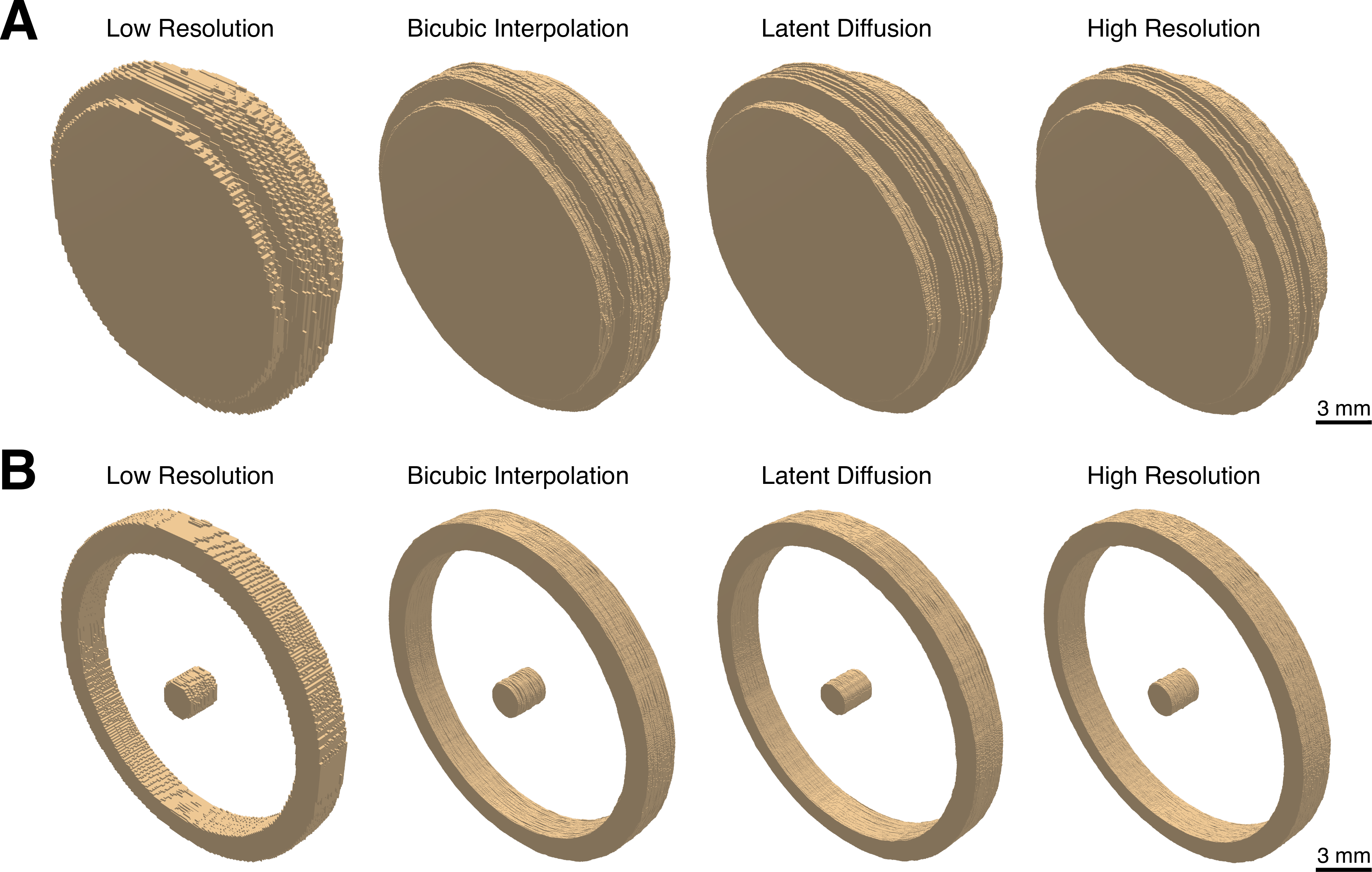}

\caption{3D samples produced from segmented  images of the build plate. These samples are shown for the low resolution webcam segmentation, the bicubic interpolation of the webcam segmentation, the model prediction based on the webcam images, and the high-resolution Basler images. \textbf{a} 3D samples visualized from Dataset A. \textbf{b)} 3D samples visualized from Dataset B.}
\label{fig:3D_samples}
\end{figure}

The qualitative results for this experiment are shown in Figure \ref{fig:part_split_qualitative}. Similar to the patch-based training scenario, we observe that the small scale details of the part are effectively reconstructed in the latent diffusion predictions. The quantitative metrics of image quality are also examined in this experiment in Figure \ref{fig:part_split_qualitative}c).

\begin{table}[]
\centering
\begin{tabular}{@{}llcccccc@{}}
\toprule
\multicolumn{2}{c}{Configuration} & \multicolumn{3}{c}{Optical Image} & \multicolumn{3}{c}{Part Reconstruction} \\ \cmidrule(lr){1-2} \cmidrule(lr){3-5} \cmidrule(lr){6-8}
Dataset   & Comparison & MAE  $\downarrow$   & PSNR $\uparrow$  & SSIM $\uparrow$   & IoU  $\uparrow$   & $H$ $\downarrow$  & $V\downarrow$   \\ \midrule
Dataset A & HR $|$ LR    & 0.23     &  9.40    &    0.64      &     0.854  &    2.05  &   0.166    \\
Dataset A & HR $|$ SR    & \textbf{0.02  }     &   \textbf{27.12}  &    \textbf{0.76 }     &  \textbf{ 0.875}    &  \textbf{ 0.36}  &  \textbf{ 0.140 }   \\ 

\bottomrule
\end{tabular}
\caption{A comparison of the dataset metrics defining the ability of the model to perform image reconstruction and part reconstruction.}
\label{tab:comparison}
\end{table}

To further demonstrate the performance of the super-resolution model, we segment a mask from the collected optical image denoting the outline of the part. By iteratively stacking the mask obtained from each layer of the build, we obtain an estimate of the 3D structure of the part. Figure \ref{fig:3D_samples} demonstrates this for a comparison of the low-resolution samples, a bicubic upscaling of the low-resolution samples, the super-resolution samples, and the ground truth samples for each part respectively. Qualitatively, we observe that the low-resolution samples and bicubic-upsampling cases have minor artifacts that prevent them from accurately capturing the geometry of the sample. We define three additional metrics to describe this behavior quantitatively. Specifically, we compute the Hausdorff Distance ($H$) between the largest contour in each image, the voxel mismatch between three-dimensional samples ($V$), and the intersection-over-union score between the 3D sample reconstructions.

The performance summary of the part-based model is summarized in Table \ref{tab:comparison}. Specifically, we observe a 91.3\% reduction in the MAE, a 188\% increase in the PSNR, and 18.8\% increase in the SSIM when comparing the low-resolution error metrics to the model prediction error metrics for Dataset A. Similarly, we observe increases in the IoU score, and corresponding decreases in the Hausdorff distance and voxel mismatch when comparing the low-resolution metrics to the model prediction metrics. 

\subsection*{Surface Roughness}

\begin{figure}[h!]

\centering
\includegraphics[width=1\linewidth]{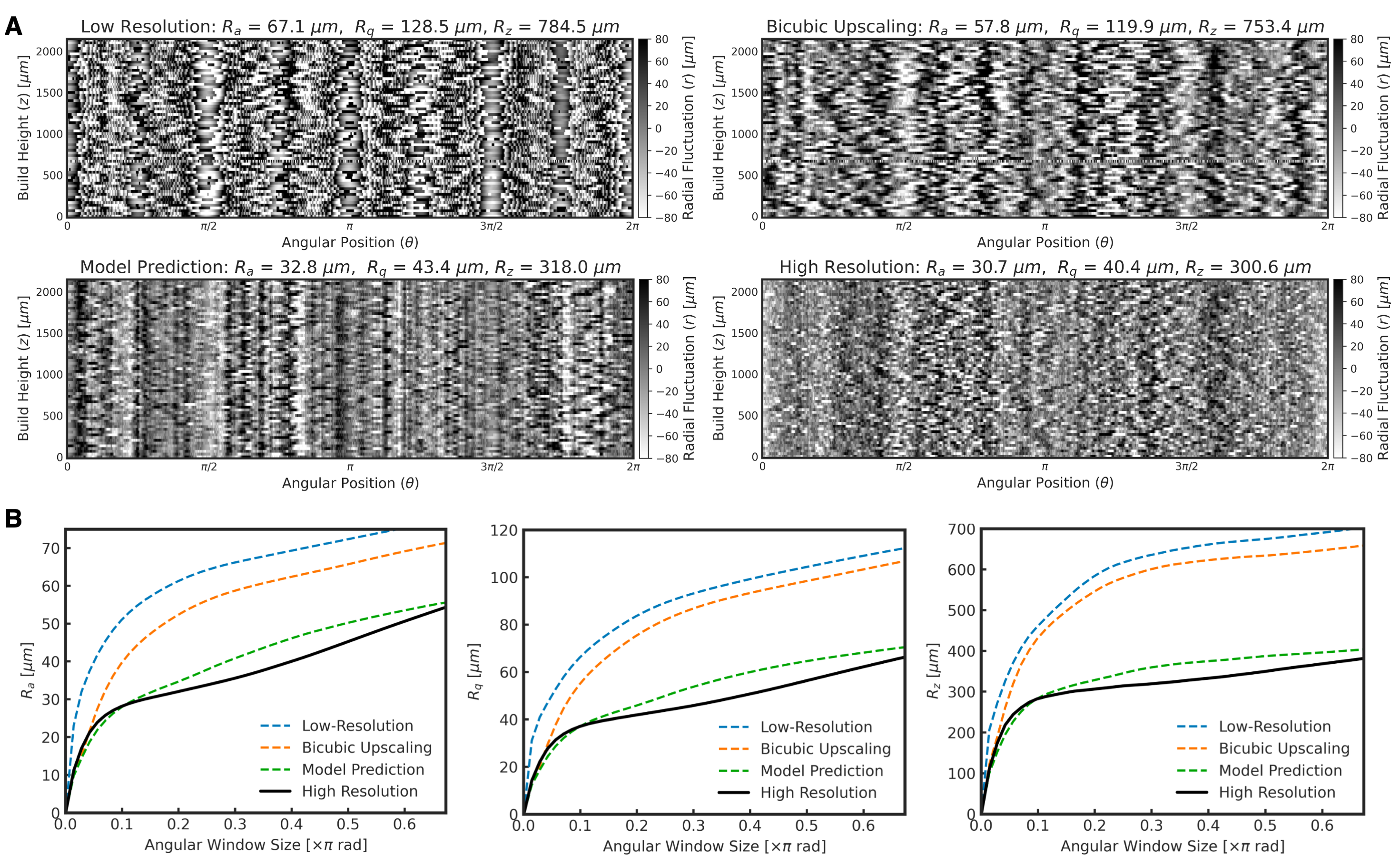}

\caption{Surface roughness as a function of the window size used for detecting the part structure for the mean-absolute error ($R_a$), RMSE ($R_q$) and maximum error ($R_z$) for two sample parts on the build plate. The bright illumination of the first build plate enables accurate surface roughness metrics, while the darker area where the second part is located provides a slightly lower degree of improvement in the surface roughness detection.}
\label{fig:surface_roughness}

\end{figure}
In order to further benchmark the model performance for evaluating part quality, we examine the ability of the latent diffusion framework to recreate accurate surface roughness profiles from the segmented model predictions. In this process, we extract the contour at each layer of the segmented part mask, and compute a line profile by projecting the contour into polar coordinates, $(r, \theta)$. The deviation from the net shape of the part for each line profile is computed by first computing a low-pass filter at a designated window size to extract the underlying part structure, before subtracting the raw signal from the low-pass filtered signal. The resulting fluctuation signal is mean-centered and aggregated for each layer to create a 2-D surface height map $\mathbf{z}( r, \theta)$. From this height map, we compute three surface roughness metrics, $R_a$, $R_q$, and $R_z$, defined in Equations \ref{eq:ra}, \ref{eq:rq}, and \ref{eq:rz} respectively. 

\begin{equation}
\label{eq:ra}
R_a = \frac{1}{N} \sum_{i=1}^{N} \left| z_i \right|
\end{equation}

\begin{equation}
\label{eq:rq}
R_q = \sqrt{\frac{1}{N} \sum_{i=1}^{N} z_i^2}
\end{equation}

\begin{equation}
\label{eq:rz}
R_z = \frac{1}{n} \sum_{j=1}^{n} \left( z_{p_j} - z_{v_j} \right)
\end{equation}

The surface roughness values extracted are plotted as a function of the window size in Figure \ref{fig:surface_roughness} to account for the absence of an exact ground-truth net shape, and to account for minor continuous deviations from a ideal circular cross-section due to the skew and warping process. These comparisons are made visually for a sample in Dataset A contained entirely within the test partition of the dataset. For this sample shown, the surface roughness extracted from the composited model predictions aligns more closely with the surface roughness extracted from the high-resolution samples. This is shown quantitatively by examining the surface roughness metrics, $R_a$, where $R_a$ = 30.7 $\mu m$ for the high-resolution sample, $R_a$ = 32.8 $\mu m$ for the latent diffusion predictions, and $R_a$ = 57.8 $\mu m$ for the bicubically interpolated low-resolution data.

\subsection*{Generalization Considerations}

\begin{figure}[h!]

\centering
\includegraphics[width=1\linewidth]{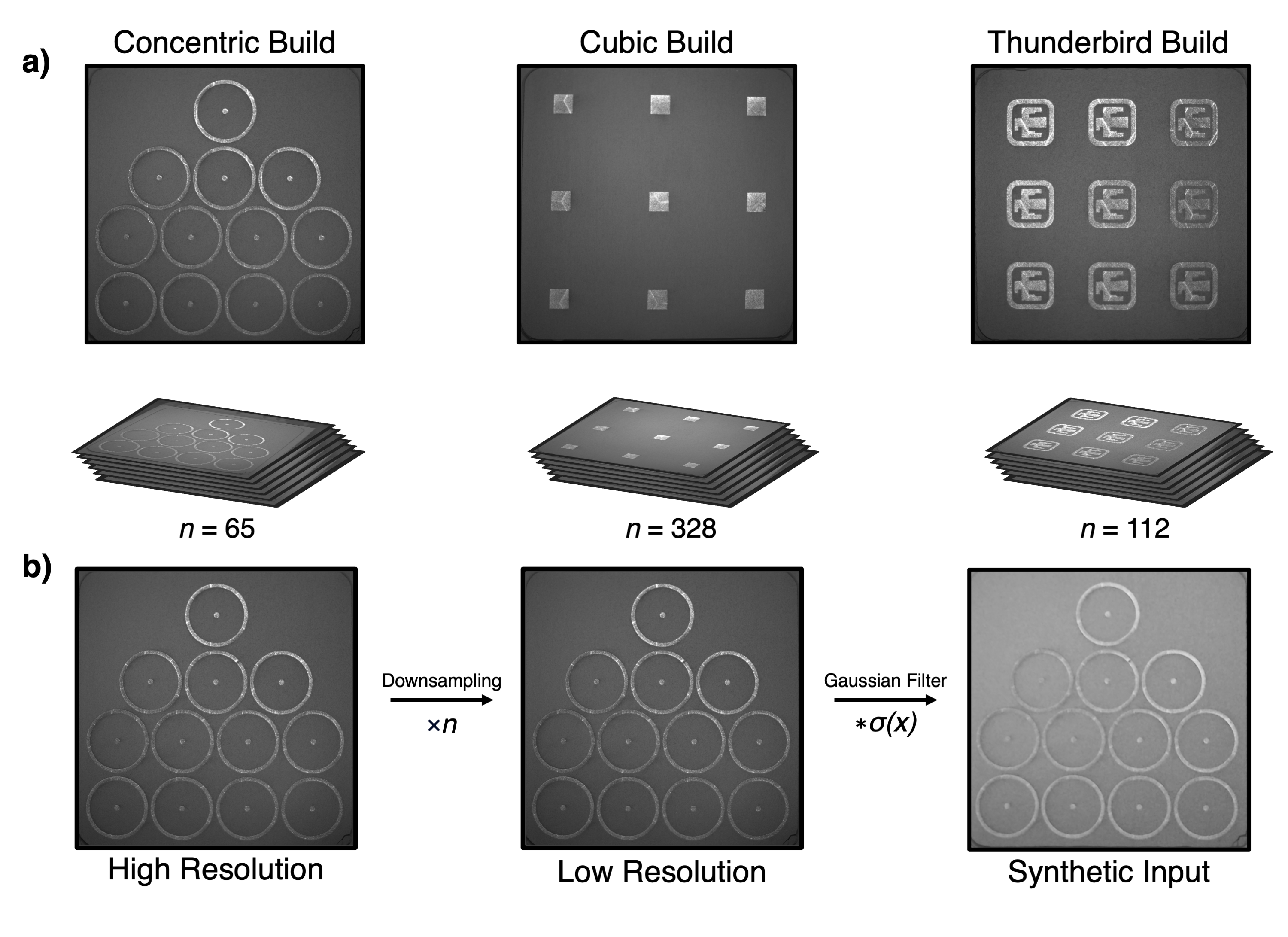}

\caption{\textbf{a)} Layerwise images of two additional parts are added to the dataset, where the cubic build has 328 layers available and the thunderbird build has 112 layers available for training. \textbf{b)} Synthetic low-resolution data is created by downsampling by a scale factor of $n=5.1$ and applying a Gaussian filter of kernel size $\sigma$.}
\label{fig:generalization_qualitative}

\end{figure}

\begin{figure}[htbp!]

\centering
\includegraphics[width=1\linewidth]{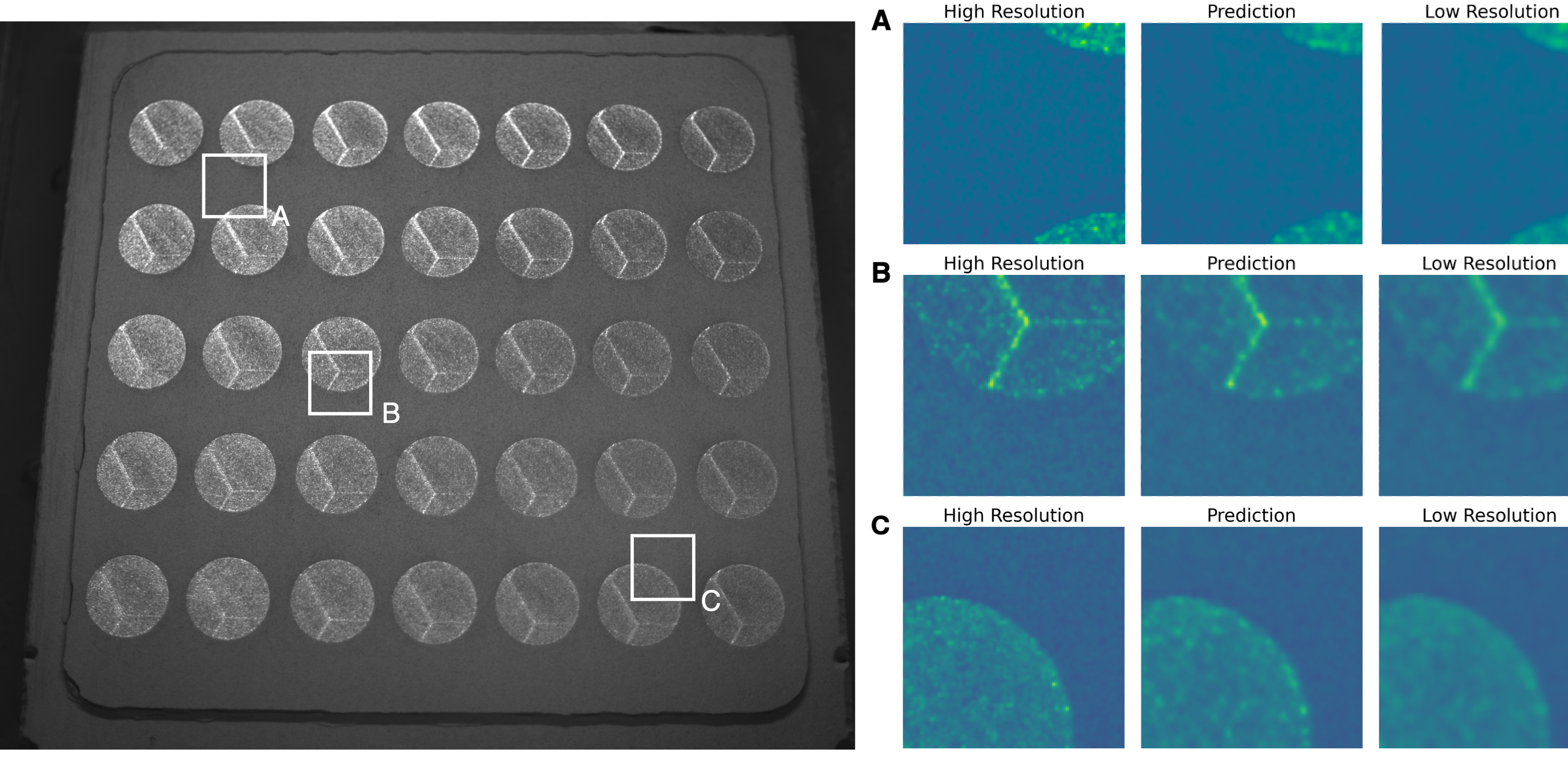}

\caption{Qualitative examples of the latent diffusion model performance on unseen parts, following synthetic downsampling at three different locations on the build plate. The Gaussian kernel size used for degradation is $\sigma$ = 10, and the magnitude of the randomly sampled noise is $\epsilon$ = 0.01.}
\label{fig:generalization_figure_qualitative}

\end{figure}

To evaluate the generalization performance of our model, we perform a series of experiments aimed at transferring a model trained on one set of sample parts to another set of sample parts. To facilitate this, we expand the dataset available to include other part cross-sections. Specifically, the initial dataset of high-resolution samples described in Figure \ref{fig:dataset_details} is augmented with two additional part shapes (Figure \ref{fig:generalization_qualitative}). For several of these part builds, only high-resolution Basler camera data is available. Therefore, we synthetically construct corresponding low-resolution images from the high-resolution images such that the degradation present within the low-resolution images matches that present in the webcam feed. Specifically, we first downsample the high-resolution image by a factor \textit{n}, where \textit{n}   is computed from the ratio of the low-resolution and high-resolution images (\textit{n = 5.1}). Following this downsampling step, a Gaussian filter is applied with kernel size $\sigma$ (Figure \ref{fig:generalization_qualitative}b).

\begin{table}[]
\centering
\begin{tabular}{@{}lccccc@{}}
\toprule
\multicolumn{2}{c}{Configuration} & \multicolumn{4}{c}{Zero-shot Metrics}
\\ \cmidrule(lr){1-2} \cmidrule(lr){3-6}
Dataset   & Model & MAE  $\downarrow$   & PSNR $\uparrow$  & SSIM $\uparrow$   & nCVD  $\downarrow$  \\ \midrule
Hold-out & $n=1$, $ \sigma=10$    &  0.029 $\pm$ 0.024   &  21.9 $\pm$ 4.59 &   0.62 $\pm$ 0.15 & $4.40 \pm  1.64 $     \\
Hold-out  &$n=4$, $ \sigma=10$   & 0.014 $\pm$ 0.007 & 26.5 $\pm$  3.49 &   0.69 $\pm$ 0.12 &  $4.50 \pm 1.65$ \\ \midrule
\bottomrule
\end{tabular}
\caption{A comparison of the dataset metrics as a function of the kernel size of the Gaussian filter used for degradation ($\sigma$), and the number of parts used for training ($n$).}
\label{tab:patch_comparison_generalization}
\end{table}

Leveraging this synthetic dataset, we extract smaller patches of size 64 $\times$ 64 from the synthetic input data and train a latent diffusion model to upscale these images to the high-resolution target. The model configuration is identical to the configuration described above. However, a train-test split is designed to enable the model to be evaluated on a part cross-section not encountered during training. Specifically, one entire part build is omitted from the dataset and hereafter referred to as the \textit{hold-out} dataset partition. The model is trained on the three other build examples shown in Figure \ref{fig:generalization_qualitative}. With this partitioning split we can evaluate the performance of the model on both inter-layer and inter-part generalization tasks. Following the training process, we first benchmark the qualitative agreement between the model predictions and the high-resolution ground truth. Multiple configurations of the model training process are defined by the number of individual build that are used following the synthetic data process to train the model, ($n$), in addition to the kernel size of the Gaussian filter ($\sigma$).

We note the similarity of our model performance in both inter-layer and inter-part generalization tasks. This is shown qualitatively in Figure \ref{fig:generalization_qualitative} for three patch samples on the unseen build plate to demonstrate inter-part generalization. In each patch, we observe the reconstruction of the localized bright areas within the sample prediction, despite the model not having encountered similar structures within the training set. We next seek to compare the inter-part generalization performance with the capability of the model to perform inter-layer generalization tasks. The performance of the model has been demonstrated in previous sections on the inter-layer generalization task, therefore, this analysis serves as a method to evaluate any performance degradation present when moving to another part cross-section. The result of this analysis is shown in Table \ref{tab:patch_comparison_generalization}, for various model configurations varying in $n$ and $\sigma$. While the large-scale metrics, such as the PSNR, SSIM and MAE exhibit relatively little variation, the covariance distance metric demonstrates the effectiveness of training a model on multiple parts prior to zero-shot analysis. Specifically, the covariance distance decreases to 2.93 $\times$ $10^5$ from 2.21 $\times$ $10^6$ as $n$ increases from $n = 1$ to $n = 4$.

\begin{figure}[htbp!]

\centering
\includegraphics[width=1\linewidth]{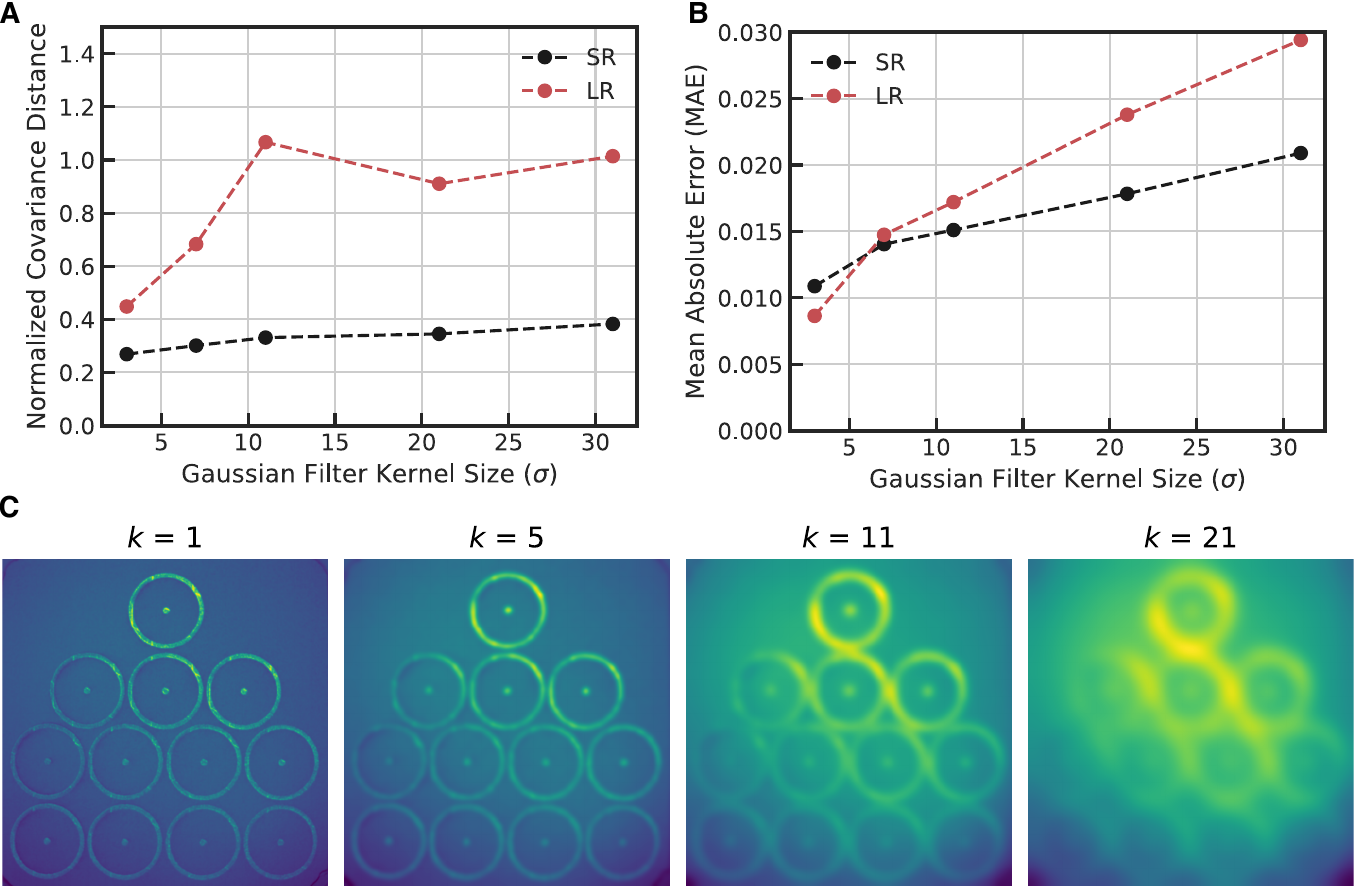}

\caption{\textbf{a)} The normalized covariance distance between the high-resolution and low-resolution (LR) samples, compared to the normalized covariance distance between the high-resolution and model predictions (SR) as a function of the Gaussian kernel filter. \textbf{b)} The mean absolute error (MAE) between the high-resolution and low-resolution (LR) samples, compared to the SSIM between the high-resolution and model predictions (SR) as a function of the Gaussian kernel filter. c) A comparison of the optical image quality as the size of the Gaussian filter kernel increases. }
\label{fig:SSIM_Norm_SR}

\end{figure}

We also seek to evaluate the performance of the latent diffusion framework in reconstructing information from lower resolution optical images than those captured by the webcam used here. To do this, we design an experiment where the Gaussian kernel size used to filter the high-resolution data is varied from small values (3) that preserve both high and low-frequency information, to large values (31) that obscure high-frequency information. A separate low-resolution encoder and latent diffusion model are trained on each kernel size configuration, with the model hyper-parameters held constant to those reported in Table \ref{tab:config_params_ae} and Table \ref{tab:config_params_ld}. Regions of size $128 \times 128$ are extracted from the layer-wise optical image for the training and inference process. The reconstruction results are shown alongside  sample synthetic low-resolution build plates in Figure \ref{fig:SSIM_Norm_SR}. We compare the normalized covariance distance and the MAE metrics as a function of the Gaussian kernel size. As the kernel size increases and the low-resolution data degrades in quality, the normalized covariance distance and MAE increase accordingly. However, the applied latent diffusion model consistently retains a stable covariance distance as the kernel size increases. This demonstrates that the texture of the predicted image remains consistent with the high-resolution image even at significant levels of degradation. The MAE increases as a function of the kernel size for both the model predictions  and the synthetic low-resolution data. However, the rate of increase is significantly larger for the low-resolution data. Therefore, the implemented latent diffusion model is able to recover detail in both the image texture and the large-scale image structure.

\FloatBarrier
\section*{Conclusion}
\label{sec:Conclusion}
In this work, we present a framework for probabilistically upsampling high-resolution optical layer-wise images from low-resolution, low-fidelity webcam images. This framework implements a latent diffusion pipeline to efficiently perform prediction with a reduced sampling time compared to a pixel-space diffusion process. Specifically, an autoencoder network is first trained to encode the high-resolution and low-resolution data into individual latent spaces. Next, a diffusion model is trained in the space of the latent vectors to reconstruct a corresponding high-resolution image from a low-resolution input. The performance of this framework is evaluated and compared to deterministic methods, in both the preservation of the image quality metrics, and the agreement between the three-dimensional reconstructed shapes. By examining these extracted metrics, we observe an increased PSNR and SSIM and significantly decreased MAE in the space of image metrics, and corresponding improvements in performance in the examined three-dimensional metrics. Additionally, by comparing the model's ability to capture image texture through complex wavelet transforms, we observe significant decreases in the defined surface texture metric between the low-resolution and high-resolution samples. Finally, we demonstrate the ability of the model to accurately preserve the surface roughness of the reconstructed samples. The development of this framework represents a step towards enabling in-situ optical monitoring at high-resolutions minimizing the trade-off between detail and monitoring location area. Future directions of this work may include incorporating this framework into an end-to-end monitoring system for detecting process events. Additionally, this work could also be extended through correlations of the optical imaging data to process conditions and porosity formation. 

\section*{Acknowledgements}

Sandia National Laboratories is a multimission laboratory managed and operated
by National Technology \& Engineering Solutions of Sandia, LLC (NTESS), a wholly owned
subsidiary of Honeywell International Inc., for the U.S. Department of Energy’s
National Nuclear Security Administration (DOE/NNSA) under contract
DE-NA0003525.
This written work is authored by an employee of NTESS. The employee, not NTESS, owns the right, title and interest in and to the written work and is responsible for its contents. Any subjective views or opinions that might be expressed in the written work do not necessarily represent the views of the U.S. Government. The publisher acknowledges that the U.S. Government retains a non-exclusive, paid-up, irrevocable, world-wide license to publish or reproduce the published form of this written work or allow others to do so, for U.S. Government purposes. The DOE will provide public access to results of federally sponsored research in accordance with the DOE Public Access Plan.

 \bibliographystyle{elsarticle-num} 
 \bibliography{cas-refs}

\begin{thebibliography}{10}
\expandafter\ifx\csname url\endcsname\relax
  \def\url#1{\texttt{#1}}\fi
\expandafter\ifx\csname urlprefix\endcsname\relax\def\urlprefix{URL }\fi
\expandafter\ifx\csname href\endcsname\relax
  \def\href#1#2{#2} \def\path#1{#1}\fi

\bibitem{li2020review}
Y.~Li, Z.~Feng, L.~Hao, L.~Huang, C.~Xin, Y.~Wang, E.~Bilotti, K.~Essa,
  H.~Zhang, Z.~Li, et~al., A review on functionally graded materials and
  structures via additive manufacturing: from multi-scale design to versatile
  functional properties, Advanced Materials Technologies 5~(6) (2020) 1900981.

\bibitem{reeves2011additive}
P.~Reeves, C.~Tuck, R.~Hague, Additive manufacturing for mass customization,
  in: Mass customization, Springer, 2011, pp. 275--289.

\bibitem{beaman2020additive}
J.~Beaman, D.~L. Bourell, C.~Seepersad, D.~Kovar, Additive manufacturing
  review: early past to current practice, Journal of Manufacturing Science and
  Engineering 142~(11) (2020) 110812.

\bibitem{Beese2016}
A.~M. Beese, B.~E. Carroll,
  \href{https://link.springer.com/article/10.1007/s11837-015-1759-z}{Review of
  mechanical properties of ti-6al-4v made by laser-based additive manufacturing
  using powder feedstock}, JOM 68 (2016) 724--734.
\newblock \href {https://doi.org/10.1007} {\path{doi:10.1007}}.
\newline\urlprefix\url{https://link.springer.com/article/10.1007/s11837-015-1759-z}

\bibitem{Tofail2018}
S.~A. Tofail, E.~P. Koumoulos, A.~Bandyopadhyay, S.~Bose, L.~O'Donoghue,
  C.~Charitidis, Additive manufacturing: scientific and technological
  challenges, market uptake and opportunities, Materials Today 21 (2018)
  22--37.
\newblock \href {https://doi.org/10.1016/J.MATTOD.2017.07.001}
  {\path{doi:10.1016/J.MATTOD.2017.07.001}}.

\bibitem{debroy2018additive}
T.~DebRoy, H.~Wei, J.~Zuback, T.~Mukherjee, J.~Elmer, J.~Milewski, A.~M. Beese,
  A.~d. Wilson-Heid, A.~De, W.~Zhang, Additive manufacturing of metallic
  components--process, structure and properties, Progress in Materials Science
  92 (2018) 112--224.

\bibitem{Morgan2004}
R.~Morgan, C.~J. Sutcliffe, W.~O'Neill,
  \href{https://link.springer.com/article/10.1023/B:JMSC.0000013875.62536.fa}{Density
  analysis of direct metal laser re-melted 316l stainless steel cubic
  primitives}, Journal of Materials Science 39 (2004) 1195--1205.
\newblock \href {https://doi.org/10.1023/B:JMSC.0000013875.62536.FA/METRICS}
  {\path{doi:10.1023/B:JMSC.0000013875.62536.FA/METRICS}}.
\newline\urlprefix\url{https://link.springer.com/article/10.1023/B:JMSC.0000013875.62536.fa}

\bibitem{Jia2014}
Q.~Jia, D.~Gu, Selective laser melting additive manufacturing of inconel 718
  superalloy parts: Densification, microstructure and properties, Journal of
  Alloys and Compounds 585 (2014) 713--721.
\newblock \href {https://doi.org/10.1016/J.JALLCOM.2013.09.171}
  {\path{doi:10.1016/J.JALLCOM.2013.09.171}}.

\bibitem{Carlton2016}
H.~D. Carlton, A.~Haboub, G.~F. Gallegos, D.~Y. Parkinson, A.~A. MacDowell,
  Damage evolution and failure mechanisms in additively manufactured stainless
  steel, Materials Science and Engineering: A 651 (2016) 406--414.
\newblock \href {https://doi.org/10.1016/J.MSEA.2015.10.073}
  {\path{doi:10.1016/J.MSEA.2015.10.073}}.

\bibitem{Sames2014}
W.~J. Sames, F.~Medina, W.~H. Peter, S.~S. Babu, R.~R. Dehoff,
  \href{https://onlinelibrary.wiley.com/doi/full/10.1002/9781119016854.ch32}{Effect
  of process control and powder quality on inconel 718 produced using electron
  beam melting}, 8th International Symposium on Superalloy 718 and Derivatives
  2014 (2014) 409--423\href {https://doi.org/10.1002/9781119016854.CH32}
  {\path{doi:10.1002/9781119016854.CH32}}.
\newline\urlprefix\url{https://onlinelibrary.wiley.com/doi/full/10.1002/9781119016854.ch32}

\bibitem{Darvish2016}
K.~Darvish, Z.~W. Chen, T.~Pasang, Reducing lack of fusion during selective
  laser melting of cocrmo alloy: Effect of laser power on geometrical features
  of tracks, Materials\& Design 112 (2016) 357--366.
\newblock \href {https://doi.org/10.1016/J.MATDES.2016.09.086}
  {\path{doi:10.1016/J.MATDES.2016.09.086}}.

\bibitem{Rehman2021}
A.~U. Rehman, M.~A. Mahmood, F.~Pitir, M.~U. Salamci, A.~C. Popescu, I.~N.
  Mihailescu, \href{https://www.mdpi.com/2079-4991/11/12/3284/htm
  https://www.mdpi.com/2079-4991/11/12/3284}{Keyhole formation by laser
  drilling in laser powder bed fusion of ti6al4v biomedical alloy: Mesoscopic
  computational fluid dynamics simulation versus mathematical modelling using
  empirical validation}, Nanomaterials 2021, Vol. 11, Page 3284 11 (2021) 3284.
\newblock \href {https://doi.org/10.3390/NANO11123284}
  {\path{doi:10.3390/NANO11123284}}.
\newline\urlprefix\url{https://www.mdpi.com/2079-4991/11/12/3284/htm
  https://www.mdpi.com/2079-4991/11/12/3284}

\bibitem{cunningham2019keyhole}
R.~Cunningham, C.~Zhao, N.~Parab, C.~Kantzos, J.~Pauza, K.~Fezzaa, T.~Sun,
  A.~D. Rollett, Keyhole threshold and morphology in laser melting revealed by
  ultrahigh-speed x-ray imaging, Science 363~(6429) (2019) 849--852.

\bibitem{foster2015optical}
B.~Foster, E.~Reutzel, A.~Nassar, B.~Hall, S.~Brown, C.~Dickman, Optical,
  layerwise monitoring of powder bed fusion (2015).

\bibitem{davna2019influence}
M.~Da{\v{n}}a, I.~Zetkov{\'a}, P.~Hanzl, The influence of a ceramic recoater
  blade on 3d printing using direct metal laser sintering, Manufacturing
  Technology 19~(1) (2019) 23--28.

\bibitem{Kayacan2019}
M.~Y. Kayacan, K.~Özsoy, B.~Duman, N.~Yilmaz, M.~C. Kayacan,
  \href{https://www.tandfonline.com/doi/abs/10.1080/10426914.2019.1655151}{A
  study on elimination of failures resulting from layering and internal
  stresses in powder bed fusion (pbf) additive manufacturing}, Materials and
  Manufacturing Processes 34 (2019) 1467--1475.
\newblock \href {https://doi.org/10.1080/10426914.2019.1655151}
  {\path{doi:10.1080/10426914.2019.1655151}}.
\newline\urlprefix\url{https://www.tandfonline.com/doi/abs/10.1080/10426914.2019.1655151}

\bibitem{smith2016spatially}
R.~J. Smith, M.~Hirsch, R.~Patel, W.~Li, A.~T. Clare, S.~D. Sharples, Spatially
  resolved acoustic spectroscopy for selective laser melting, Journal of
  Materials Processing Technology 236 (2016) 93--102.

\bibitem{yadroitsev2014selective}
I.~Yadroitsev, P.~Krakhmalev, I.~Yadroitsava, Selective laser melting of
  ti6al4v alloy for biomedical applications: Temperature monitoring and
  microstructural evolution, Journal of alloys and compounds 583 (2014)
  404--409.

\bibitem{berumen2010quality}
S.~Berumen, F.~Bechmann, S.~Lindner, J.-P. Kruth, T.~Craeghs, Quality control
  of laser-and powder bed-based additive manufacturing (am) technologies,
  Physics procedia 5 (2010) 617--622.

\bibitem{myers2023high}
A.~J. Myers, G.~Quirarte, F.~Ogoke, B.~M. Lane, S.~Z. Uddin, A.~B. Farimani,
  J.~L. Beuth, J.~A. Malen, High-resolution melt pool thermal imaging for
  metals additive manufacturing using the two-color method with a color camera,
  Additive Manufacturing (2023) 103663.

\bibitem{pak2024thermopore}
P.~M.-W. Pak, F.~Ogoke, A.~Polonsky, A.~Garland, D.~S. Bolintineanu, D.~R.
  Moser, M.~J. Heiden, A.~B. Farimani, Thermopore: Predicting part porosity
  based on thermal images using deep learning, arXiv preprint arXiv:2404.16882
  (2024).

\bibitem{abdelrahman2017flaw}
M.~Abdelrahman, E.~W. Reutzel, A.~R. Nassar, T.~L. Starr, Flaw detection in
  powder bed fusion using optical imaging, Additive Manufacturing 15 (2017)
  1--11.

\bibitem{li2018situ}
Z.~Li, X.~Liu, S.~Wen, P.~He, K.~Zhong, Q.~Wei, Y.~Shi, S.~Liu, In situ 3d
  monitoring of geometric signatures in the powder-bed-fusion additive
  manufacturing process via vision sensing methods, Sensors 18~(4) (2018) 1180.

\bibitem{Mohr2020}
G.~Mohr, S.~J. Altenburg, A.~Ulbricht, P.~Heinrich, D.~Baum, C.~Maierhofer,
  K.~Hilgenberg, \href{https://www.mdpi.com/2075-4701/10/1/103/htm
  https://www.mdpi.com/2075-4701/10/1/103}{In-situ defect detection in laser
  powder bed fusion by using thermography and optical tomography—comparison
  to computed tomography}, Metals 2020, Vol. 10, Page 103 10 (2020) 103.
\newblock \href {https://doi.org/10.3390/MET10010103}
  {\path{doi:10.3390/MET10010103}}.
\newline\urlprefix\url{https://www.mdpi.com/2075-4701/10/1/103/htm
  https://www.mdpi.com/2075-4701/10/1/103}

\bibitem{pagani2020automated}
L.~Pagani, M.~Grasso, P.~J. Scott, B.~M. Colosimo, Automated layerwise
  detection of geometrical distortions in laser powder bed fusion, Additive
  Manufacturing 36 (2020) 101435.

\bibitem{zur2013high}
J.~Zur~Jacobsm{\"u}hlen, S.~Kleszczynski, D.~Schneider, G.~Witt, High
  resolution imaging for inspection of laser beam melting systems, in: 2013
  IEEE international instrumentation and measurement technology conference
  (I2MTC), IEEE, 2013, pp. 707--712.

\bibitem{ashby2022thermal}
A.~Ashby, G.~Guss, R.~K. Ganeriwala, A.~A. Martin, P.~J. DePond, D.~J. Deane,
  M.~J. Matthews, C.~L. Druzgalski, Thermal history and high-speed optical
  imaging of overhang structures during laser powder bed fusion: A
  computational and experimental analysis, Additive Manufacturing 53 (2022)
  102669.

\bibitem{caltanissetta2018characterization}
F.~Caltanissetta, M.~Grasso, S.~Petr{\`o}, B.~M. Colosimo, Characterization of
  in-situ measurements based on layerwise imaging in laser powder bed fusion,
  Additive Manufacturing 24 (2018) 183--199.

\bibitem{imani2018process}
F.~Imani, A.~Gaikwad, M.~Montazeri, P.~Rao, H.~Yang, E.~Reutzel, Process
  mapping and in-process monitoring of porosity in laser powder bed fusion
  using layerwise optical imaging, Journal of Manufacturing Science and
  Engineering 140~(10) (2018) 101009.

\bibitem{gobert2018application}
C.~Gobert, E.~W. Reutzel, J.~Petrich, A.~R. Nassar, S.~Phoha, Application of
  supervised machine learning for defect detection during metallic powder bed
  fusion additive manufacturing using high resolution imaging., Additive
  Manufacturing 21 (2018) 517--528.

\bibitem{SNOW202112}
Z.~Snow, B.~Diehl, E.~W. Reutzel, A.~Nassar,
  \href{https://www.sciencedirect.com/science/article/pii/S027861252100008X}{Toward
  in-situ flaw detection in laser powder bed fusion additive manufacturing
  through layerwise imagery and machine learning}, Journal of Manufacturing
  Systems 59 (2021) 12--26.
\newblock \href {https://doi.org/https://doi.org/10.1016/j.jmsy.2021.01.008}
  {\path{doi:https://doi.org/10.1016/j.jmsy.2021.01.008}}.
\newline\urlprefix\url{https://www.sciencedirect.com/science/article/pii/S027861252100008X}

\bibitem{boschetto2024powder}
A.~Boschetto, L.~Bottini, S.~Vatanparast, Powder bed monitoring via digital
  image analysis in additive manufacturing, Journal of Intelligent
  Manufacturing 35~(3) (2024) 991--1011.

\bibitem{li2022srdiff}
H.~Li, Y.~Yang, M.~Chang, S.~Chen, H.~Feng, Z.~Xu, Q.~Li, Y.~Chen, Srdiff:
  Single image super-resolution with diffusion probabilistic models,
  Neurocomputing 479 (2022) 47--59.

\bibitem{wang2018esrgan}
X.~Wang, K.~Yu, S.~Wu, J.~Gu, Y.~Liu, C.~Dong, Y.~Qiao, C.~Change~Loy, Esrgan:
  Enhanced super-resolution generative adversarial networks, in: Proceedings of
  the European conference on computer vision (ECCV) workshops, 2018, pp. 0--0.

\bibitem{yang2019deep}
W.~Yang, X.~Zhang, Y.~Tian, W.~Wang, J.-H. Xue, Q.~Liao, Deep learning for
  single image super-resolution: A brief review, IEEE Transactions on
  Multimedia 21~(12) (2019) 3106--3121.

\bibitem{zhang2018residual}
Y.~Zhang, Y.~Tian, Y.~Kong, B.~Zhong, Y.~Fu, Residual dense network for image
  super-resolution, in: Proceedings of the IEEE conference on computer vision
  and pattern recognition, 2018, pp. 2472--2481.

\bibitem{kim2016accurate}
J.~Kim, J.~K. Lee, K.~M. Lee, Accurate image super-resolution using very deep
  convolutional networks, in: Proceedings of the IEEE conference on computer
  vision and pattern recognition, 2016, pp. 1646--1654.

\bibitem{goodfellow2014generative}
I.~Goodfellow, J.~Pouget-Abadie, M.~Mirza, B.~Xu, D.~Warde-Farley, S.~Ozair,
  A.~Courville, Y.~Bengio, Generative adversarial nets, Advances in neural
  information processing systems 27 (2014).

\bibitem{gayon2020pores}
A.~Gayon-Lombardo, L.~Mosser, N.~P. Brandon, S.~J. Cooper, Pores for thought:
  generative adversarial networks for stochastic reconstruction of 3d
  multi-phase electrode microstructures with periodic boundaries, npj
  Computational Materials 6~(1) (2020) 82.

\bibitem{ogoke2022deep}
O.~F. Ogoke, K.~Johnson, M.~Glinsky, C.~Laursen, S.~Kramer, A.~B. Farimani,
  Deep-learned generators of porosity distributions produced during metal
  additive manufacturing, Additive Manufacturing 60 (2022) 103250.

\bibitem{rakotonirina2020esrgan+}
N.~C. Rakotonirina, A.~Rasoanaivo, Esrgan+: Further improving enhanced
  super-resolution generative adversarial network, in: ICASSP 2020-2020 IEEE
  International Conference on Acoustics, Speech and Signal Processing (ICASSP),
  IEEE, 2020, pp. 3637--3641.

\bibitem{gao2023implicit}
S.~Gao, X.~Liu, B.~Zeng, S.~Xu, Y.~Li, X.~Luo, J.~Liu, X.~Zhen, B.~Zhang,
  Implicit diffusion models for continuous super-resolution, in: Proceedings of
  the IEEE/CVF conference on computer vision and pattern recognition, 2023, pp.
  10021--10030.

\bibitem{thanh2020catastrophic}
H.~Thanh-Tung, T.~Tran, Catastrophic forgetting and mode collapse in gans, in:
  2020 international joint conference on neural networks (ijcnn), IEEE, 2020,
  pp. 1--10.

\bibitem{liu2019spectral}
K.~Liu, W.~Tang, F.~Zhou, G.~Qiu, Spectral regularization for combating mode
  collapse in gans, in: Proceedings of the IEEE/CVF international conference on
  computer vision, 2019, pp. 6382--6390.

\bibitem{srivastava2017veegan}
A.~Srivastava, L.~Valkov, C.~Russell, M.~U. Gutmann, C.~Sutton, Veegan:
  Reducing mode collapse in gans using implicit variational learning, Advances
  in neural information processing systems 30 (2017).

\bibitem{ho2020denoising}
J.~Ho, A.~Jain, P.~Abbeel, Denoising diffusion probabilistic models, Advances
  in Neural Information Processing Systems 33 (2020) 6840--6851.

\bibitem{ogoke2024inexpensive}
F.~Ogoke, Q.~Liu, O.~Ajenifujah, A.~Myers, G.~Quirarte, J.~Malen, J.~Beuth,
  A.~B. Farimani, Inexpensive high fidelity melt pool models in additive
  manufacturing using generative deep diffusion, Materials \& Design (2024)
  113181.

\bibitem{song2020denoising}
J.~Song, C.~Meng, S.~Ermon, Denoising diffusion implicit models, arXiv preprint
  arXiv:2010.02502 (2020).

\bibitem{Rombach_2022_CVPR}
R.~Rombach, A.~Blattmann, D.~Lorenz, P.~Esser, B.~Ommer, High-resolution image
  synthesis with latent diffusion models, in: Proceedings of the IEEE/CVF
  Conference on Computer Vision and Pattern Recognition (CVPR), 2022, pp.
  10684--10695.

\bibitem{ronneberger2015u}
O.~Ronneberger, P.~Fischer, T.~Brox, U-net: Convolutional networks for
  biomedical image segmentation, in: Medical image computing and
  computer-assisted intervention--MICCAI 2015: 18th international conference,
  Munich, Germany, October 5-9, 2015, proceedings, part III 18, Springer, 2015,
  pp. 234--241.

\bibitem{kirillov2023segment}
A.~Kirillov, E.~Mintun, N.~Ravi, H.~Mao, C.~Rolland, L.~Gustafson, T.~Xiao,
  S.~Whitehead, A.~C. Berg, W.-Y. Lo, et~al., Segment anything, in: Proceedings
  of the IEEE/CVF International Conference on Computer Vision, 2023, pp.
  4015--4026.

\bibitem{wang2004image}
Z.~Wang, A.~C. Bovik, H.~R. Sheikh, E.~P. Simoncelli, Image quality assessment:
  from error visibility to structural similarity, IEEE transactions on image
  processing 13~(4) (2004) 600--612.

\bibitem{zhang2021maximum}
S.~Zhang, S.~Mallat, Maximum entropy models from phase harmonic covariances,
  Applied and Computational Harmonic Analysis 53 (2021) 199--230.

\bibitem{grossmann1990reading}
A.~Grossmann, R.~Kronland-Martinet, J.~Morlet, Reading and understanding
  continuous wavelet transforms, in: Wavelets: Time-Frequency Methods and Phase
  Space Proceedings of the International Conference, Marseille, France,
  December 14--18, 1987, Springer, 1990, pp. 2--20.

\bibitem{esser2021taming}
P.~Esser, R.~Rombach, B.~Ommer, Taming transformers for high-resolution image
  synthesis, in: Proceedings of the IEEE/CVF conference on computer vision and
  pattern recognition, 2021, pp. 12873--12883.

\end{thebibliography}

\appendix

\section{Wavelet Covariance Metric}
\label{sec:wavelets}

The wavelet covariance metric provides a measure of the difference between the coherent structures present within a pair of images. Here, it is used to calculate the similarity between the ground truth, low-resolution, and model-predicted images in terms of the high-frequency details present within the image. This metric is defined by the difference in the phase harmonic covariance of each image, shown in Equation \ref{eq:loss} and reproduced below in Equation \ref{eq:loss_appendix}. In Equation \ref{eq:loss_appendix},  $\tilde{K}_{\mathcal{R}x}$ is the covariance of the wavelet coefficients produced by the ground truth image, and $ \tilde{K}_{\mathcal{R}\bar{x}}$ is the covariance of the wavelet coefficients produced by the trial image.

\begin{equation}
\label{eq:loss_appendix}
    f(x) = \| \tilde{K}_{\mathcal{R}x} - \tilde{K}_{\mathcal{R}\bar{x}} \|
\end{equation}

The reduced-order phase harmonic representation $\mathcal{R}(x)$ of a signal $x$ is computed by applying a phase harmonic operator $\hat{\mathcal{H}}$ to the output of a complex wavelet transform $\mathcal{W}x$ (Equation \ref{eq:rx}). 

\begin{equation}
\label{eq:rx}
\mathcal{R}x = \hat{\mathcal{H}}(\mathcal{W}x) = \left \{ \hat{h}(k) [ x \star \psi_\lambda (u) ]^k \right \}
\end{equation}

The complex wavelet transform convolves $x$ with a set of complex wavelets $\psi_\lambda (\omega)$, defined for varying rotation, dilation, and translation operations. The index $\lambda = (j, r)$ combines the set of dilation at scales $j$ and rotation operations at angles $r$ applied to the wavelet, while $u$ indexes the set of translation operations applied to the wavelet basis. Wavelets are computed up to a maximum scale $j = J$, where any lower frequencies are captured by a “father" wavelet which acts as a low-pass filter. The convolution operations are performed by computing point-wise multiplication between the Fourier transform of the wavelet $\hat{\psi}$ and the Fourier transform of the input signal $\hat{x}$.

\begin{equation}
\hat{\phi}(\omega) = \exp \left ( -  \frac{\left | \omega \right |^2}{2 \sigma^2} \right )
\end{equation}

The complex bump steerable wavelet is used to perform the convolution operation. This wavelet is defined in the Fourier domain by Equation \ref{eq:bump_steerable}. Complex steerable wavelets are introduced to examine the properties of images at multiple scales and orientations, where the property of steerability ensures that the wavelets are rotated copies of each other and a given wavelet can be computed through linear combinations of the basis wavelets. 
\begin{equation}
\label{eq:bump_steerable}
\hat{\psi}(\omega) = c \cdot \exp \left( -\frac{(|\omega| - \xi_0)^2}{\xi_0^2 - (|\omega| - \xi_0)^2} \right) \mathbf{1}_{[0,2\xi_0]}(|\omega|) \cdot \cos^{Q/2-1} (\text{arg}(\omega)) \mathbf{1}_{\text{arg}(\omega) < \frac{\pi}{2}}
\end{equation}

Following the convolution operation, a phase harmonic operator is applied to the wavelet coefficients to introduce correlations across scales. A phase harmonic of the signal is calculated by exponentiating the phase component only by an integer $k$, preserving the modulus (Equation \ref{eq:phase_harmonic_operator}).

\begin{equation}
\label{eq:phase_harmonic_operator}
[z]^{k} = \left | z \right | e^{ik\varphi(z)}
\end{equation}

The phase harmonic operator, $\hat{\mathcal{H}}$ computes all such phase harmonics of the signal, and weights them by a harmonic weight $\hat{h}(k)$. These phase harmonic weights modify the contribution of individual phase harmonics and tend to attenuate high values of $k$. This operation effectively acts as a phase filter, similar to the non-linear activation function used within neural networks. The phase harmonic weights used are derived from the phase window $\cos (\alpha) > 0$, where $\alpha$ is the phase variable. The corresponding harmonic weights are derived from this expression by computing the Fourier integral, and are reproduced below in Equation \ref{eq:fourier_weights}.

\begin{equation}
\label{eq:fourier_weights}
\hat{h}(k) =
\begin{cases}
    \frac{(-1)^{k/2 + 1}}{\pi (k^2 - 1)} & \text{if } k \text{ is even} \\
    \frac{1}{4} & \text{if } k = \pm 1 \\
    0 & \text{if } |k| > 1 \text{ is odd}
\end{cases}
\end{equation}

\begin{figure}[htbp!]

\centering
\includegraphics[width=1\linewidth]{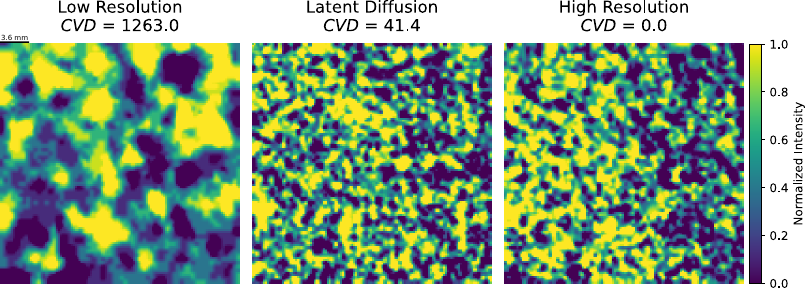}

\caption{A comparison of the powder bed textures for the low-resolution, latent model diffusion prediciton and high-resolution images. The model prediction correctly recreates the scale of the intensity variation, which is reflected in the decrease in the covariance distance metric.}
\label{fig:texture_distance_demonstration}

\end{figure}

An example of the information present within the texture metric is shown in Figure \ref{fig:texture_distance_demonstration}. The patches shown here are taken for a sample comprised only of the powder bed, without the part component visible. The low-resolution is extracted from the webcam feed, while the high-resolution image is extracted from the Basler camera feed. The high-resolution and low-resolution images demonstrate markedly different scales of variation. This is reflected in the relative covariance distances compared between the HR and SR image, and the HR and LR images. The SR image reconstructs the correct scale of variability for the powder bed sample, and achieves a distance of 41.4 compared to the LR distance of 1263.0.

\section{Autoencoder Performance Details}
We train two autoencoder networks to facilitate the latent diffusion process. As in \cite{Rombach_2022_CVPR} and \cite{esser2021taming}, we use a two-dimensional convolutional encoder and decoder to create the latent space. As diffusion models are parameterized for optimal performance on data with coordinate unit variance \cite{ho2020denoising}, a KL-penalty between the standard normal distribution and the learned latent space is applied. Furthermore, the latent space is rescaled prior to the diffusion process by the estimated component-wise variance \cite{Rombach_2022_CVPR}. The loss function used to train the autoencoder consists of a reconstruction loss term, the KL-divergence term, an adversarial loss term, and a perceptual loss term. The loss function is reproduced in Equation \ref{eq:autoencoder_loss} from \cite{esser2021taming, Rombach_2022_CVPR}. 
\begin{equation}
\label{eq:autoencoder_loss}
L_{\text{Autoencoder}} = \min_{\mathcal{E}, \mathcal{D}} \max_{\psi} \left( L_{\text{rec}}(x, \mathcal{D}(\mathcal{E}(x))) - L_{\text{adv}}(\mathcal{D}(\mathcal{E}(x))) + \log D_{\psi}(x) + L_{\text{reg}}(x; \mathcal{E}, \mathcal{D}) \right)
\end{equation}

One autoencoder is designed to compress the low-resolution image into a conditioning vector, while the second autoencoder is used to compress the dimension used for the diffusion process. For a successful latent diffusion process, it is important that the two encoder networks are able to accurately reconstruct the images as they appear in the ground truth image. 

\begin{figure}[htbp!]

\centering
\includegraphics[width=1\linewidth]{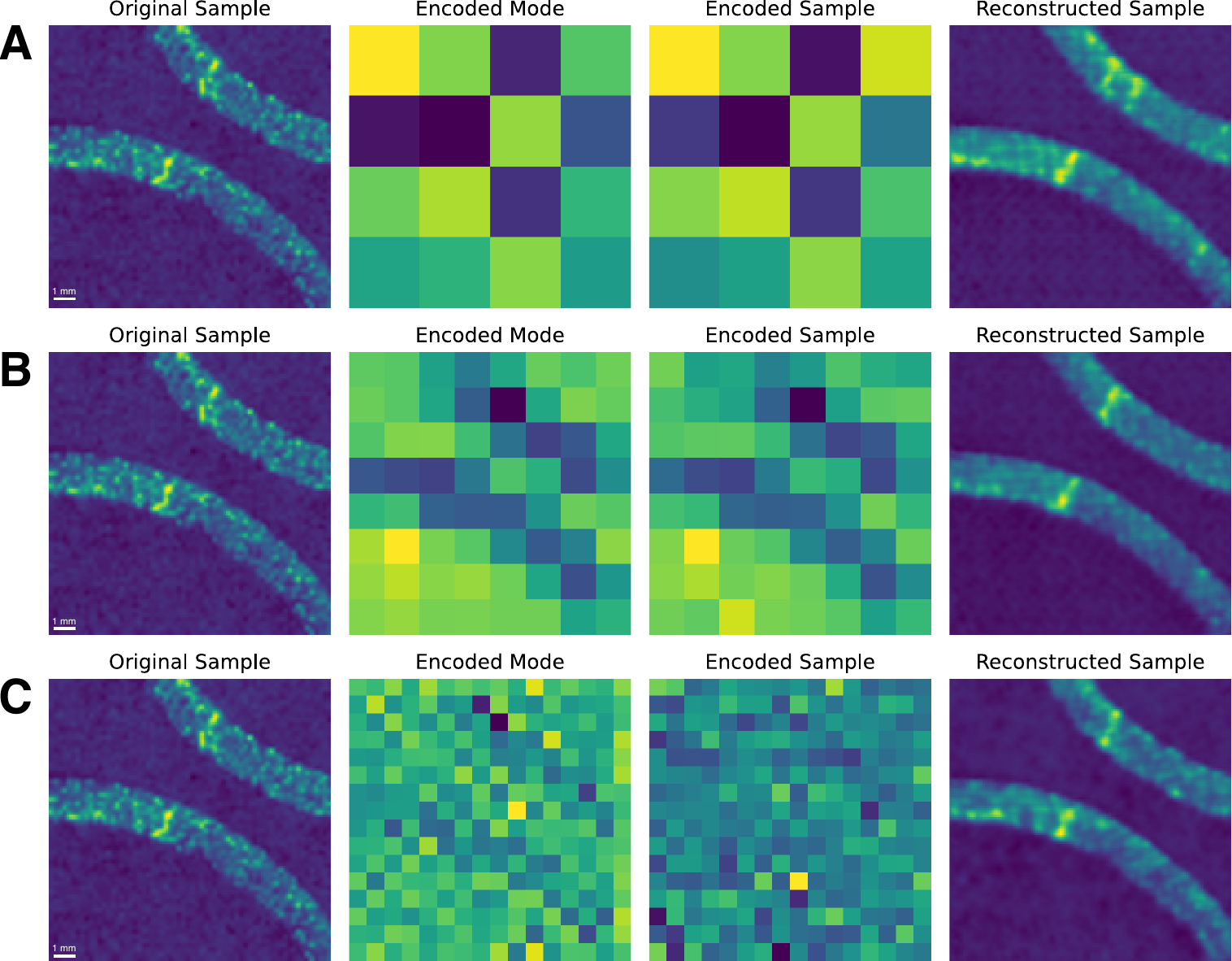}

\caption{Visualizations of a 64$\times$64 pixel region from the high resolution optical image, the first channel of the mode of the variational autoencoder latent vector, the first channel of a sample from the variational autoencoder latent vector, and the sample reconstruction at varying latent space sizes. \textbf{a)} Visualizations corresponding to a 4x4 latent space. \textbf{b)} Visualizations corresponding to an 8$\times$8 latent space. \textbf{c)} Visualizations corresponding to a 16$\times$16 latent space.}
\label{fig:aecomparison}

\end{figure}

We evaluate the effect of varying the compression dimension on the image reconstruction quality in Table \ref{tab:autoencoder_compression} and Figure \ref{fig:aecomparison}. As the size of the latent space increases, the fidelity of the reconstruction improves, reducing the amount of artifacts present in the image.

\begin{table}[htbp!]
\centering
\begin{tabular}{@{}lccccc@{}}
\toprule
\multicolumn{2}{c}{Configuration} & \multicolumn{4}{c}{Autoencoder Reconstruction}
\\ \cmidrule(lr){1-2} \cmidrule(lr){3-5}
Dataset   & Latent Space & MAE  $\downarrow$   & PSNR $\uparrow$  & SSIM $\uparrow$    $\downarrow$  \\ \midrule
Dataset A & 4 $\times$4     &  0.0133   &  32.4 &   0.928   \\
Dataset A & 8 $\times 8$    & 0.0127   &  32.4 &   0.936  \\
Dataset A & 16 $\times$16     &  0.009   &  36.0 &   0.966   \\\midrule
\bottomrule
\end{tabular}
\caption{A comparison of the image evaluation metrics as a function of the latent space size used for autoencoder compression.}
\label{tab:autoencoder_compression}
\end{table}
\FloatBarrier
\section{Model Architecture Details}

The hyperparameters used for the variational autoencoder and latent diffusion models are reported in Table \ref{tab:config_params_ae} and  Table \ref{tab:config_params_ld} respectively.
\begin{table}[htbp!]
    \centering
    \caption{Autoencoder Model Parameters}
    \label{tab:config_params_ae}
    \begin{tabular}{ll}
        \toprule
        \textbf{Model} & \textbf{Parameters} \\
        \midrule
        Learning Rate &  $4.5\times 10^{-6}$  \\
    
        KL-divergence Weight & $1\times 10^{-6}$ \\
        Discriminator Loss Weight & 0.5 \\
        
        Latent Space Channels & 4 \\
        Latent Space Feature Map & 16 $\times$ 16 \\

        \bottomrule
    \end{tabular}
\end{table}

\begin{table}[htbp!]
    \centering
    \caption{Diffusion U-Net Parameters}
    \label{tab:config_params_ld}
    \begin{tabular}{ll}
        \toprule
        \textbf{Model} & \textbf{Parameters} \\
        \midrule
        Learning Rate & 5$\times 10^{-6}$ \\
        Latent Channels & 4 \\
        Resolution & 128 \\
        Timesteps & 200 \\

         Baseline Model Channels & 128 \\
       Channel Multiplication Factor & \{1, 2, 4, 4\} \\
       Residual Blocks & 2 \\

        \bottomrule
    \end{tabular}
\end{table}






\end{document}